\documentclass[11pt,a4paper]{article}
\usepackage[utf8]{inputenc}
\usepackage{amsmath}
\usepackage{amsfonts}
\usepackage{amssymb}
\usepackage{bm}
\usepackage{graphicx}
\usepackage[left=2cm,right=2cm,top=2cm,bottom=2cm]{geometry}
\usepackage{multirow}
\usepackage{authblk}
\usepackage{enumitem}
\usepackage{setspace}
\usepackage{bbm}
\usepackage{chngcntr}
\usepackage{url}
\usepackage{makecell}
\usepackage{rotating}
\usepackage{booktabs}

\title{\textbf{Optimal weighted tests for replication studies and the two-trials rule}}

\author[a]{David S.\ Robertson\thanks{Corresponding author \texttt{david.robertson@mrc-bsu.cam.ac.uk}}}
\author[a,b]{Thomas Jaki}

\affil[a]{MRC Biostatistics Unit, University of Cambridge, UK}
\affil[b]{University of Regensburg, Germany}

\date{\vspace{-24pt}}

\begin{document}

\maketitle

\onehalfspacing

\begin{abstract}
Replication studies for scientific research are an important part of ensuring the reliability and integrity of experimental findings. In the context of clinical trials, the concept of replication has been formalised by the `two-trials' rule, where two pivotal studies are required to show positive results before a drug can be approved. In experiments testing multiple hypotheses simultaneously, control of the overall familywise error rate (FWER) is additionally required in many contexts. The well-known Bonferroni procedure controls the FWER, and a natural extension is to introduce weights into this procedure to reflect the \textit{a-priori} importance of hypotheses or to maximise some measure of the overall power of the experiment. In this paper, we consider analysing a replication study using an optimal weighted Bonferroni procedure, with the weights based on the results of the original study that is being replicated and the optimality criterion being to maximise the disjunctive power of the trial (the power to reject at least one non-null hypothesis). We show that using the proposed procedure can lead to a substantial increase in the disjunctive power of the replication study, and is  robust to changes in the effect sizes between the two studies. \\

\end{abstract}

\section{Introduction}


Driven by long-standing concerns of a reproducibility crisis in science~\cite{Ioannidis2005}, the importance of replicating scientific studies is well-recognised~\cite{Jasny2011}. Such replication studies are now commonplace in all areas of scientific research. In the context of clinical trials, for example, this has been formalised by the US Food and Drug Administration (FDA) in what is known as the \textit{two-trials rule}. This states that there should be ``at least two pivotal studies, each convincing on its own"~\cite{FDA1998} before a drug is approved and ``reflects the need for substantiation of experimental results, which has often been referred to as the need for replication of the finding''~\cite{FDA2019}. For trials with a single hypothesis, the two-trial rule is usually satisfied by running two independent trials and requiring the results to be statistically significant at the standard (one-sided) 2.5\% level~\cite{Senn2021, Held2024}.

In trials (and scientific experiments more generally) that test multiple hypotheses simultaneously, there is the added complication of how to handle this multiplicity. In confirmatory (or `pivotal') trials, where the aim is to provide definitive results, the \textit{familywise error rate} (FWER) is the typical error rate to be controlled~\cite{EMA2017, FDA2022}. The FWER is defined as the (maximum) probability of falsely rejecting at least one null hypotheses, under any configuration of which hypotheses are null and non-null. Control of the FWER at some small level $\alpha$ is seen as particularly important when the hypotheses are related in some way, such as when testing multiple doses of the same drug or multiple endpoints on the same group of patients.

Many multiple testing procedures have been proposed to control the FWER in clinical trials and other scientific contexts more generally~\cite{Dudoit2008}. A popular choice (and perhaps the most common by far) remains the well-known Bonferroni procedure. When testing~$m$ hypotheses of interest $H_1, \ldots, H_m$ with corresponding $p$-values denoted $p_1, \ldots, p_m$, the Bonferroni procedure rejects hypothesis $H_i$ if $p_i < \alpha/m$. A natural extension is to consider a \textit{weighted} Bonferroni procedure, where instead of `splitting' the allowed type~I error ($\alpha$) equally between the hypotheses, pre-determined weights $w_i$ are assigned to hypothesis~$H_i$ (where these weights are non-negative and sum up to 1). Hypothesis~$H_i$ is then rejected if $p_i < w_i \alpha$. Note that setting $w_i \equiv 1/m$ recovers the standard (unweighted) Bonferroni procedure. Intuitively, weighting increases the power to reject hypothesis $H_i$ when $w_i > 1/m$, and decreases the power when $w_i < 1/m$, compared with using the standard Bonferroni procedure. These weights could be chosen based on the \textit{a-priori} relative importance of the hypotheses, or could be chosen to \textit{optimise} different definitions of the power of the experiment. The latter is the approach we consider in this paper. 

The idea of choosing the weights to optimise power for Bonferroni procedures has received limited attention in the literature. Almost all of this has focused on maximising either the mean power, the weighted mean power (where the weights reflect their \textit{a-priori} importance), or the expected number of rejections. In the context of fixed-sequence tests for clinical trials, optimal weights have been provided numerically to maximise the weighted power~\cite{Westfall2001} and the expected number of rejections~\cite{Zhang2015}. In three related papers~\cite{Wasserman2006, Rubin2006, Roeder2009} in the context of replication studies for genomics, analytical formulae have been derived for weighted Bonferroni tests to obtain the optimal mean power.

As discussed in~\cite{Westfall2001, Xi2024}, while the mean (weighted or unweighted) power is easier to optimise numerically, it has issues with interpretability and is not the most common definition of overall power used. For this reason, Xi and Chen~\cite{Xi2024} considered choosing the weights to maximise either the \textit{disjunctive} or \textit{conjunctive} power of a single trial with multiple hypotheses. The disjunctive power is the probability of rejecting at least one non-null hypothesis, whereas the conjunctive power is the probability of rejecting all non-null hypotheses~\cite{Senn2007}. The motivation for using the disjunctive power in the context of a clinical trial would be the desire to find at least one truly effective treatment, or demonstrate the effectiveness of a treatment on at least one of the (co-)primary endpoints of the trial.  Xi and Chen~\cite{Xi2024} showed how to maximise either the disjunctive or conjunctive power for both independent and correlated test statistics, with the optimal weights found numerically. 

A key challenge in all of the above literature is that the true optimal weights depend on the true effect sizes (i.e.\ treatment effects in the clinical trial setting) associated with each hypothesis, which are unknown. In~\cite{Westfall2001, Zhang2015, Xi2024} this problem is handled by using \textit{a-priori} effect sizes for each hypothesis, corresponding to the effect sizes required for the desired marginal power and sample size calculations (for example). Meanwhile, in~\cite{Wasserman2006, Rubin2006, Roeder2009} a sample-splitting approach was proposed, where part of the data from the experiment is used to estimate the true effect size, and the other part of the data is used for hypothesis testing. This latter approach can suffer from a substantial loss in power, however, due to the reduced sample size.

In this paper, we take a different approach for the context of replication studies and the two trial rule in particular. Given that an experiment (labelled 'experiment 2') is designed as a replication study for some earlier experiment (labelled `experiment 1'), we can use the data from experiment~1 to estimate the true effect sizes, which are then used to calculate optimal weights for testing the hypotheses in experiment~2. Like in Xi and Chen~\cite{Xi2024}, we consider maximising the disjunctive power as our optimality criteria, but propose an alternative numerical solution to the optimisation problem. We also consider the robustness of our proposal, in terms of what happens when the effect sizes in experiments~1 and~2 differ.



The rest of the paper proceeds as follows. Section~\ref{sec:theory} develops the theoretical framework for the optimal weighted Bonferroni procedure to maximise disjunctive power. In Section~\ref{sec:case_study} we present a real-life case study of a `two-trial' submission to the FDA with multiple treatment groups and endpoints. We present a comprehensive simulation study in Section~\ref{sec:sim_study} and conclude with a Discussion in Section~\ref{sec:discuss}.

\section{Methods}
\label{sec:theory}


Suppose we have $m>1$ hypotheses of interest denoted $H_1, \ldots, H_m$, where (without loss of generality) these hypotheses are of the form $H_i : \theta_i = 0$ versus the alternatives $H_i': \theta_i > 0$ for $i = 1, \ldots, m$. In the clinical trial context, $\theta_i$ could correspond to the treatment effect for experimental treatment~$i$, or alternatively the treatment effect for a single experimental treatment as measured by endpoint~$i$ (in a trial with multiple endpoints). We let $\bm{\theta} = (\theta_1, \ldots, \theta_m)$ denote the vector of true parameter values and $\mathcal{A} = \{ i  : \theta_i > 0 \}$ denote the index set of alternative (i.e., non-null) hypotheses.

Two independent trials/experiments are conducted to test this set of multiple hypotheses, resulting in two independent sets of standardised test statistics $\bm{T^{(j)}} = (T_1^{(j)}, \ldots, T_m^{(j)})$ for $j = 1, 2$, where $T_i^{(j)} \sim N(\theta_i, 1)$, at least asymptotically. The labelling $j=1$ corresponds to the trial/experiment that finishes first. The resulting $p$-values from the two experiments are denoted $p_1^{(j)}, \ldots, p_m^{(j)}$ for $j = 1, 2$, where $p_i^{(j)} = \bar{\Phi}(T_i^{(j)})$, $\bar{\Phi}(\cdot) = 1 - \Phi(\cdot)$ and $\Phi(\cdot)$ denotes the standard normal cdf (cumulative distribution function).

\subsection{Weighted Bonferroni procedure}

In order to control the FWER for each experiment, we use the (standard) Bonferroni procedure for experiment~1 and a \textit{weighted} Bonferroni procedure for experiment~2, where these weights depend on the data from experiment~1. Hence, for experiment~1, hypothesis $H_i$ is rejected if $p_i^{(1)} < \alpha/m$. For experiment~2, hypothesis $H_i$ is rejected if $p_i^{(2)} < w_i \alpha$, where $w_i$ is the pre-determined weight assigned to hypothesis $H_i$. Note that a weighted Bonferroni procedure could also be used for experiment~1 given some \textit{a-priori} weights, reflecting the relative importance of the hypotheses for example, but for simplicity we only consider using the usual unweighted procedure. More generally, \textit{any} multiple hypothesis testing approach that controls the FWER could be used for experiment~1.


Given this formulation, we wish to use the observed test statistics from experiment~1 to choose \textit{optimal} weights to use for experiment~2. The optimality criterion we consider in this paper is to maximise the \textit{disjunctive power}, i.e. the probability to reject at least one non-null hypothesis. Other definitions of power could be used, such as the average power or the conjunctive power, which we return to in the Discussion. 

In what follows, we make the simplifying assumption that the test statistics in experiment~2 are all independent of one another. This can happen, for example, if each test statistic corresponds to data collected from distinct groups of subjects/patients. We return to the issue of correlated test statistics in the Discussion. The \textit{marginal} probability to reject hypothesis $H_i$ using the weighted Bonferroni procedure in experiment~2 is 
\[
P(p_i^{(2)} < w_i  \alpha ) = P(T_i^{(2)} > \bar{\Phi}^{-1} ( w_i \alpha )) = \bar{\Phi} \! \left( \bar{\Phi}^{-1} (w_i \alpha  ) - \theta_i \right)
\] where $\bar{\Phi}^{-1}(x) = \Phi^{-1} (1 - x)$.
Assuming independence between the $p$-values (or, equivalently, the test statistics), then the disjunctive power in experiment~2 is given by the following expression \[
1 - \prod_{i \in \mathcal{A}} P(p_i^{(2)} > w_i \alpha  )   = 1 - \prod_{i \in \mathcal{A}} \Phi \! \left( \bar{\Phi}^{-1}(w_i \alpha ) - \theta_i \right).
\]

We can then formulate the optimisation problem (as in~\cite{Xi2024}) as follows:
\begin{align}
& \max_{w_i} \qquad \quad 1 - \prod_{i \in \mathcal{A}} \Phi \! \left( \bar{\Phi}^{-1}(w_i \alpha ) - \theta_i \right) \nonumber  \\ 
& \text{subject to} \quad \sum_{i=1}^m w_i = 1 \label{eq:optim_problem}\\
& \phantom{subject to} \quad \; w_i \geq 0 \quad \text{for } i = 1, \ldots, m \nonumber
\end{align}

Given values for $\theta_1, \ldots, \theta_m$ (we return to this point later), this optimisation problem can be solved using numerical non-linear optimisation algorithms to find the set of optimal Bonferroni weights, as detailed in~\cite{Xi2024}. However, particularly as the number of hypotheses~$m$ increases, common implementations of such algorithms such as those provided by the \texttt{nloptr} R package can often run into computational issues (i.e., converging to a local rather than a global maximum). Xi and Chen~\cite{Xi2024} propose resolving these through the use of an exponentially increasing (in $|\mathcal{A}|$) number of starting points for the numerical optimisation algorithm.

In this paper, we propose an alternative solution in order to remove the need to use multiple starting points. It can be shown (see the Appendix, Section~\ref{Asec:proof} for a proof) that given values for $\theta_1, \ldots, \theta_m$, the optimal Bonferroni weights can be written in the following form: \begin{equation} \label{eq:optim_weights}
w_i = \mathbbm{1} (i \in \mathcal{A})\,  \frac{1}{\alpha} \, \bar{\Phi} \! \! \left( \frac{\theta_i}{2} + \frac{1}{\theta_i} \! \left[ \log(c) - \! \!\sum_{j \in \mathcal{A} , \, j \neq i} \log \! \left( \Phi (\bar{\Phi}^{-1} (w_j \alpha ) - \theta_j ) \right) \right] \right)
\end{equation}
where $\mathbbm{1}(\cdot)$ is the indicator function and the constant~$c>0$ is chosen to ensure that $\sum_{i=1}^m w_i = 1$. This system of non-linear simultaneous equations can be solved numerically, for example by using the \texttt{nelqslv} package in R. In Appendix~\ref{Asec:computational_methods} we compare the performance of the different computational methods. These comparisons show that the use of the \texttt{nelqslv} approach is substantially quicker than using \texttt{nloptr} with multiple starting points, and also gives higher disjunctive power (i.e.\ the \texttt{nloptr} approach can still result in suboptimal weights). For example, for $m = 5$ treatments, \texttt{nelqslv} is on average 28 times quicker than \texttt{nloptr} and always has greater or equal disjunctive power, with a strictly greater disjunctive power in almost 10\% of cases.

As an alternative to numerical solutions to the optimisation problem (either via equations~\eqref{eq:optim_problem} or equation~\eqref{eq:optim_weights}), for small~$m$ (e.g., $m = 2,3$) an exhaustive grid search can be performed. We evaluate the disjunctive power (given in equation~\eqref{eq:optim_problem}) over the space $(w_1, \ldots, w_m) \in [0, 1]^m$ subject to the constraint $\sum_{i=1}^m w_i = 1$ to find the values of $(w_1, \ldots, w_m)$ that maximise the disjunctive power. 
As shown in Appendix~\ref{Asec:computational_methods}, the \texttt{nelqslv} approach is substantially for $m > 2$ while never giving suboptimal weights in terms of disjunctive power.

\subsection{Estimating optimal weights in practice}
\label{subsec:est_optimal}

In order to use equations~\eqref{eq:optim_problem} or~\eqref{eq:optim_weights}, we need to provide values of $\theta_1, \ldots, \theta_m$ as well as the set of alternative hypotheses $\mathcal{A}$. This is where the fact that experiment~2 is designed as a replication study for some earlier experiment~1 is key. Given that experiments~1 and~2 have the same underlying true parameters $\theta_1, \ldots, \theta_m$, we can use a consistent estimator $\hat{\theta}_i$ for $\theta_i$ based on the completed data from experiment~1 as our assumed value of $\theta_i$ used in the optimisation problem and hence equations~\eqref{eq:optim_problem} or~\eqref{eq:optim_weights}. In what follows, for simplicity we only consider the usual maximum likelihood estimator (MLE). 

As for the set of alternative hypotheses $\mathcal{A}$, given that trial~1 uses the usual (unweighted) Bonferroni procedure to perform the hypothesis tests, we can define $\mathcal{A} = \{ i : p_i^{(1)} < \alpha/m \}$, i.e.\ $\mathcal{A}$ is the set of rejected hypotheses in experiment~1. In the remainder of this paper, we focus on the two-trial setting and use this to define $\mathcal{A}$, although the same pattern of results can be seen with other choices of definition of $\mathcal{A}$. For example, another option is to let $\mathcal{A} = \{ i : \hat{\theta}_i > 0\}$. 
One subtlety is what to do if $\mathcal{A}$ is empty, i.e.\ no hypotheses are rejected in experiment~1. A natural choice in this setting is to simply set all the weights equal to $1/m$ to recover the usual (unweighted) Bonferroni procedure, which we follow in this paper.


\section{Illustrative example}
\label{sec:case_study}

Consider the new drug application to the US FDA Center for Drug Evaluation and Research seeking approval for the use of Bepotastine Besilate Ophthalmic Solution (Bepreve) as an eye drop treatment for ocular itching associated with allergic conjunctivitus~\cite{Rashid2009}. In keeping with the FDA `two-trial rule', the applicant submitted two phase~III Conjunctival Allergen Challenge (CAC) studies: ISTA-BEPO-CS01 (hereafter denoted as `Trial~1') and CL-S\&E-0409071-P (hereafter denoted as `Trial~2'). Both studies were identical in design (except that trial~1 was single centre and trial~2 was multicentre), and evaluated the onset and duration of action of Bepreve 1.5\% and Beptreve 1.0\% in patients with acute allergic conjunctivitus. Trial subjects were randomised in a 1:1:1 ratio between the control, Beperve 1.0\% and Bepreve 1.5\%. There were 5~visits in a period of approximately 7~weeks, with visits~1 and~2 being for screening of eligible patients and the following~3 visits evaluating the efficacy and safety of Bepreve compared with control in alleviating the signs and symptoms of CAC-induced allergic conjunctivitis.

The primary efficacy variables were subject-evaluated ocular itching at 3 time points (3, 5 and 7 minutes post CAC) and investigator-evaluated conjunctival redness also at 3 time points (7, 15 and 20 minutes post CAC). The criteria for overall statistical significance for Bepreve 1.0\% or Bepreve 1.5\% was for statistical significance for the primary efficacy variables to be achieved at a majority (2/3) of the time points either visit 3 or 4 and additionally at visit 5. To adjust for multiplicity, the standard (unweighted) Bonferroni procedure was used for each trial, with overall level~$\alpha = 0.05$.

For simplicity, and in order to illustrate a wider set of results, we show how the weighted Bonferroni procedure could have been used to analyse the following subsets of possible analyses of the efficacy results:
\begin{itemize}
\item Visit 3, 7 min post-CAC (4 comparisons: 2 endpoints, 2 drug concentrations)
\item Visit 4, 7 min post-CAC (4 comparisons: 2 endpoints, 2 drug concentrations)
\item Visit 5, 7 min post-CAC (4 comparisons: 2 endpoints, 2 drug concentrations)
\item Conjunctival redness, Visit 3 (6 comparisons: 3 time points, 2 drug concentrations)
\item Conjunctival redness, Visit 4 (6 comparisons: 3 time points, 2 drug concentrations)
\item Conjunctival redness, Visit 5 (6 comparisons: 3 time points, 2 drug concentrations)
\end{itemize}

\noindent Table~\ref{tab:case_study} shows the observed trial~1 means and resulting optimal Bonferroni weights, as well as the `original' and `new' adjusted $p$-values (i.e., the adjusted $p$-values associated with the unweighted and weighted Bonferroni procedures, respectively).

\begin{table}[ht!]
\hspace{6cm}\textbf{Trial 1} \\[6pt]
\begin{tabular}{l c c}
\toprule 
\textbf{Analysis} & \textbf{Means} & \textbf{Optimal weights} \\
\midrule 
post-CAC Visit 3 & (3.93, 3.72, 2.22, 0.37) & (0.53, 0.47, 0, 0) \\
post-CAC Visit 4 & (4.99, 6.73, 2.50, 1.84) & (0.34, 0.60, 0.06, 0) \\
post-CAC Visit 5 & (6.48, 6.23, 4.19, 3.18) & (0.39, 0.36, 0.16, 0.08) \\
CR Visit 3       & (2.22, 1.62, 1.24, 0.37, 0.21, -0.65) & (1/6, 1/6, 1/6, 1/6, 1/6, 1/6) \\
CR Visit 4       & (2.50, 1.86, 1.62, 1.84, 1.55, 1.32) & (1, 0, 0, 0, 0, 0) \\
CR Visit 5       & (4.19, 3.93, 3.25, 3.18, 2.57, 1.69) & (0.31, 0.27, 0.17, 0.16, 0.08, 0) \\
\bottomrule 
\end{tabular}
\end{table}
\begin{table}[ht!]
\vspace{6pt} \hspace{6cm} \textbf{Trial 2} \\[6pt]
\begin{tabular}{l c c}
\toprule 
\textbf{Analysis} & \textbf{Original $\boldsymbol{p}$-values} & \textbf{Adjusted $\boldsymbol{p}$-values} \\
\midrule 
post-CAC Visit 3 & (0.000, 0.001, 0.021, 1) & (0.000, 0.001, 1, 1) \\
post-CAC Visit 4 & (0.000, 0.000, 0.002, 0.427) & (0.000, 0.000, 0.011, 1) \\
post-CAC Visit 5 & (0.000, 0.000, 0.000, 0.012) & (0.000, 0.000, 0.001, 0.038) \\
CR Visit 3       & (0.032, 0.101, 0.244, 1, 1, 1) & (0.032, 0.101, 0.244, 1, 1, 1) \\
CR Visit 4       & (0.004, 0.214, 0.616, 0.640, 1, 1) & (0.001, 1, 1, 1, 1, 1) \\
CR Visit 5       & (0.001, 0.012, 0.891, 0.019, 0.068, 1) & (0.000, 0.007, 0.874, 0.019, 0.135, 1) \\[4pt] 
\bottomrule 
\end{tabular}
\vspace{12pt}
\caption{Analyses of the efficacy results from the two pivotal trials of Bepreve for treatment for ocular itching associated with allergic conjunctivitus, where CAC = Conjunctival Allergen Challenge and CR = Conjunctival Redness. \label{tab:case_study}}
\end{table}

The set of non-zero optimal weights corresponds to the set of hypotheses which are rejected in trial~1 (except for Conjunctival redness, Visit 3 where no hypotheses are rejected and hence equal weights are used). The hypotheses with the largest trial~1 means get the largest weights as expected, meaning that it is easier to reject the hypothesis in trial~2 (as seen by the smaller adjusted $p$-values). In addition, hypotheses with trial~1 means close together will have weights that are almost equal, e.g.\ $H_1$ and $H_2$ for Visit 3, 7 min post-CAC. Another noticeable feature is that even if a hypotheses is rejected in trial~1, it may have a very low optimal weight if it has a substantially lower trial~1 mean than the other hypotheses, e.g.\ $H_3$ for Visit 4, 7 min post-CAC. In such cases, it becomes harder to subsequently reject this hypothesis in trial~2. However, overall -- in terms of disjunctive power -- one would still expect a gain given that the procedure is optimal for this criterion.

In all of the analyses presented in Table~\ref{tab:case_study}, there are no changes in the overall decision (i.e., which hypotheses are rejected in both trial~1 and trial~2). As an explicit (hypothetical) example of the overall decision changing, consider the Conjunctival Redness, Visit 3 analysis. Suppose the trial~1 mean for $H_1$ changed from 2.22 to 2.73 with corresponding $p$-value going from 0.0148 to 0.004 and we use a different overall $\alpha = 0.025$, so that $H_1$ (and only $H_1$) is rejected in trial~1 when using the usual (unweighted) Bonferroni procedure. Then the optimal weights are $(1, 0, 0, 0, 0)$ with trial~2 original adjusted $p$-values of $(0.032, 0.101, 0.244, 1, 1, 1)$ and trial~2 new adjusted $p$-values of $(0.005, 1, 1, 1, 1, 1)$. Hence, overall $H_1$ can be rejected when using the weighted Bonferroni procedure, but not the usual unweighted Bonferonni.

\section{Simulation studies}
\label{sec:sim_study}

In order to investigate the potential power gains of the proposed testing framework that uses a weighted Bonferroni test for experiment~2 compared with using an unweighted Bonferroni test, we conduct a simulation study. In what follows, for simplicity we use the setting and terminology of the `two-trial rule' for confirmatory clinical trials. Our simulation set-up is as follows.

\subsection{Set-up}

We test~$m>1$ treatment options in the first pivotal trial (trial~1), with parameter values $\bm{\theta} = (\theta_1, \ldots, \theta_m)$ corresponding to the true treatment means. The first pivotal trial yields standardised test statistics $\bm{T^{(1)}} = (T_1^{(1)}, \ldots, T_m^{(1)})$ where $T_i^{(1)} \sim N(\theta_i, 1)$. The second pivotal trial (trial~2) tests the same~$m$ treatment options, resulting in standardised test statistics $\bm{T^{(2)}} = (T_1^{(2)}, \ldots, T_m^{(2)})$ where this time $T_i^{(2)} \sim N(\theta_i', 1)$, i.e.\ the second pivotal trial has true treatment means given by  $\bm{\theta}' = (\theta_1', \ldots, \theta_m')$. We first consider the setting where $\bm{\theta}' = \bm{\theta}$ in Section~\ref{subsec:consistent_te} and investigate the robustness when $\bm{\theta}' \neq \bm{\theta}$ in Section~\ref{subsec:robustness}.

Table~\ref{tab:simul_scenarios} shows the different simulation scenarios considered for trials testing~$m$ treatment options where $m \in \{2, 3, 5\}$.
Note that for the scenario of two treatments, we start with a more comprehensive exploration of the parameter space in Section~\ref{subsec:two_treatments},  i.e.\ $\bm{\theta} = (\theta_1, \theta_2)$ where $0 < \theta_1, \theta_2, \leq 6$. In all simulation scenarios considered, we set $\alpha = 0.05$.

\begin{table}[ht!]
\centering
\begin{tabular}{l l|l}
\hline 
 \multicolumn{2}{c | }{\textbf{Scenario}} & \textbf{Treatment means} \\ 
\hline 
\multirow{3}{*}{Two treatments} &  & $\bm{\theta} = (0, \theta)$ \\ 
 & & $\bm{\theta} = (\theta/2, \theta)$ \\ 
 &  & $\bm{\theta} = (\theta, \theta)$ \\ 
\hline 
\multirow{2}{*}{Three treatments} &  & $\bm{\theta} = (0, 0, \theta)$ \\ 
 &  & $\bm{\theta} = (\theta/2, \theta, 2\theta)$ \\ 
 \hline
 \multirow{2}{*}{Five treatments} & & $\bm{\theta} = (0, 0,0,0,\theta)$ \\ 
 &  & $\bm{\theta} = (0, 0,\theta, \theta, \theta)$ \\
 &  & $\bm{\theta} = (\theta_1, \theta_2, \theta_3, \theta_4,\theta)$ where $\theta_i \sim U[0, \theta]$ independently\\ 
\hline 
\end{tabular} 

\caption{Summary of simulation scenarios. In all simulation scenarios considered $\alpha = 0.05$. \label{tab:simul_scenarios}}

\end{table}

In terms of the performance measures of interest, rather than looking at the disjunctive (or indeed marginal) power in trial~2, we instead look at the \textit{probability of success} (PoS) to reject a non-null hypothesis in both trials, reflecting the decision-making process used. Equivalently, in the more general experimental replication setting, this would correspond to replicating the same result of rejecting a null hypothesis in both experiments. More precisely, the \textit{marginal} PoS (mPoS) for a single non-null hypothesis $H_i$ is the probability of rejecting $H_i$ in both trial~1 and trial~2. The \textit{disjunctive} PoS (dPoS) is then the probability of rejecting at least one non-null hypothesis in both trial~1 and trial~2. Note that we do not show results for the FWER, since as expected from theory this is always controlled for each trial below the specified $\alpha$ level. We compare using the weighted Bonferroni procedure (as in Section~\ref{sec:theory}) in trial~2 with using the usual unweighted Bonferroni procedure.


Finally, in terms of the implementation of the weighted Bonferroni procedure, the numerical procedures used to find the optimal weights that maximise the disjunctive power are as follows:
\begin{itemize}
	\item For $m = 2, 3$ treatments we use an exhaustive grid search over $(w_1, \ldots, w_m) \in [0, 1]^m$, subject to the constraint $\sum_{i=1}^m w_i = 1$ with a grid size of 0.005. A grid search was used as it was computationally feasible and ruled out any numerical issues that can occur when using a non-linear optimisation or simultaneous equation solver.
	\item For $m > 3$, an exhaustive grid search was no longer computationally feasible. Hence, we used the R package \texttt{nelqslv} to solve the set of simultaneous equations given in equation~\eqref{eq:optim_weights} when there were more than 3 treatments from trial~1 with $p_i^{(1)} < \alpha/m$ (and used an exhaustive grid search otherwise). \\[-11pt]
\end{itemize}

\subsection{Consistent treatment effects}
\label{subsec:consistent_te}

\subsubsection{Two treatments}
\label{subsec:two_treatments}

\begin{figure}[ht!]
\includegraphics[width = 0.65\linewidth]{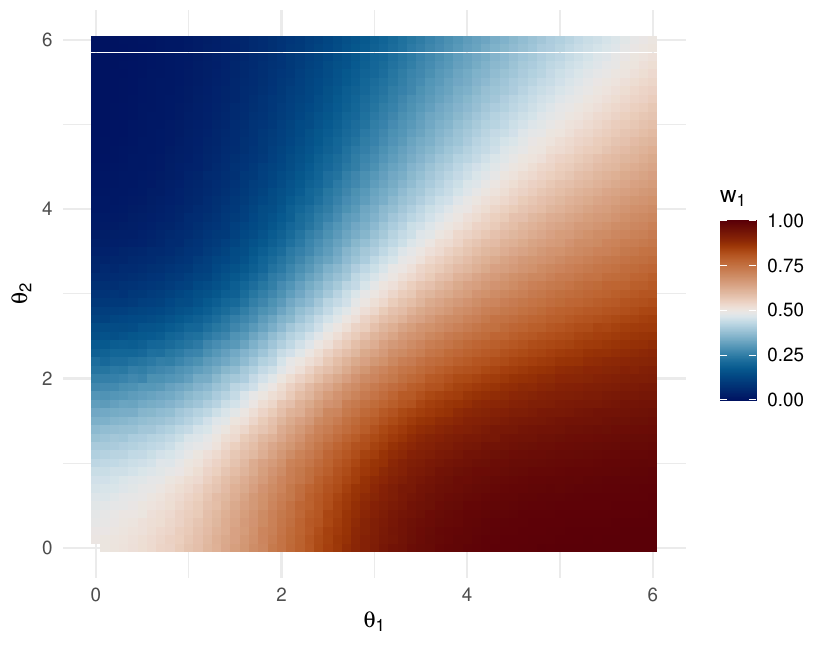} 
\centering
\caption{Heatmap showing the mean of the empirical optimal Bonferroni weights $\hat{w}_1$ across the range of parameter space of $\bm{\theta} = (\theta_1, \theta_2)$ where $0 \leq \theta_1, \theta_2 \leq 6$. Results are from $10^4$ simulations for each pair of values of $(\theta_1, \theta_2)$. \label{fig:2T_weight_heatmap}}
\end{figure}

We first explore the properties of the weighted Bonferroni approach across the parameter space of $\bm{\theta} = (\theta_1, \theta_2)$ where $0 \leq \theta_1, \theta_2 \leq 6$. Figure~\ref{fig:2T_weight_heatmap} shows the mean of the empirical weights given to hypothesis $H_1$, denoted $\hat{w}_1$, observed from $10^4$ simulation replicates for each pair of values of $(\theta_1, \theta_2)$. Note that the empirical weight given to hypothesis $H_2$, $\hat{w}_2$, is given by $\hat{w}_2 = 1 - \hat{w}_1$ by definition.

\begin{figure}[ht!]
\includegraphics[width = 0.6\linewidth]{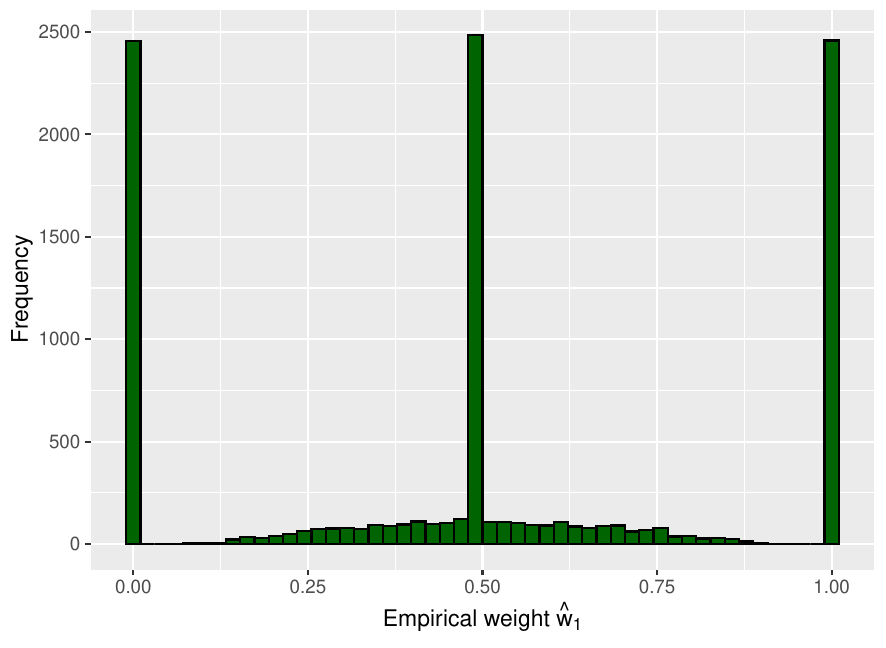} 
\centering
\caption{Histogram of the empirical optimal Bonferroni weights $\hat{w}_1$ when $\bm{\theta} = (2, 2)$. Results are from $10^4$ simulation replicates. \label{fig:2T_weight_histogram}}
\end{figure}

The mean values of $\hat{w}_1$ vary smoothly across the parameter space. As expected, when $\theta_1 = \theta_2$ then $\hat{w}_1 = \hat{w}_2 = 0.5$, i.e. the usual equal weights are recovered. We also see that $\hat{w}_1$ is an increasing function of the difference $\theta_1 - \theta_2$. When $0<\theta_2 < 1$ then $\hat{w}_1 \approx 1$ for $\theta_1 \geq 3$, and conversely when $0 < \theta_1 < 1$ then $\hat{w}_1 \approx 0$ for $\theta_2 \geq 3$. It is important to note, however, that taking the mean of $\hat{w}_1$ hides considerable stochasticity induced by the fact that the weights are based on values of $T_i^{(1)} \sim N(\theta_i, 1)$, with the PoS additionally based on the values of $T_i^{(2)} \sim N(\theta_i, 1)$.  Figure~\ref{fig:2T_weight_histogram} shows the histogram of the optimal Bonferroni weights observed from $10^4$ simulation replicates when $\bm{\theta} = (2,2)$. 

On average, the value of  $\hat{w}_1$ (and hence $\hat{w}_2$) is indeed equal to 0.5. However, the realised values of $\hat{w}_1$ (and hence $\hat{w}_2$) have three large (and approximately equal) `spikes' at 0, 0.5 and 1. These spikes at 0 and 1 correspond to trial replicates where $T_1^{(1)}$ is substantially smaller and larger than $T_2^{(1)}$, respectively.

\begin{figure}[ht!]
\includegraphics[width = 0.7\linewidth]{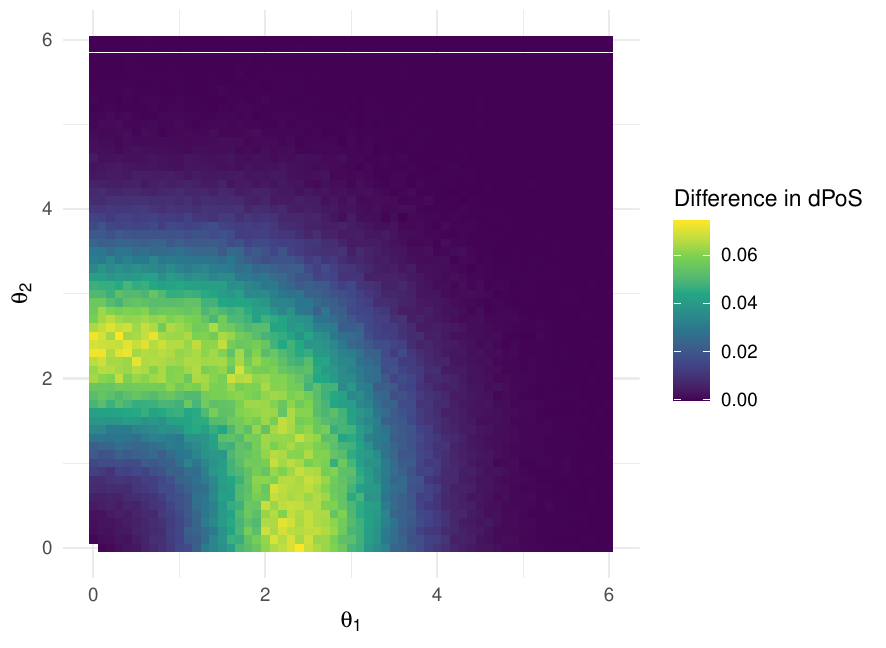} 
\centering
\caption{Heatmap showing the difference in the disjunctive Probability of Success (dPoS) across the range $\bm{\theta} = (\theta_1, \theta_2)$ where $0 \leq \theta_1, \theta_2, \leq 6$. Results are from $10^4$ simulations for each pair of values of $(\theta_1, \theta_2)$. Positive values indicate where the dPoS of the weighted Bonferroni approach is higher than the dPoS of the unweighted Bonferroni approach. \label{fig:dPoS_heatmap}}
\end{figure}

We next evaluate how these weights translate into differences in the dPoS between the weighted and unweighted Bonferroni approaches across the same parameter space. Figure~\ref{fig:dPoS_heatmap} shows a heatmap of the difference in dPoS (again based on $10^4$ simulation replicates), where a positive difference indicates where the dPoS of weighted Bonferroni is higher than the dPoS of unweighted Bonferroni.

For all regions of the parameter space, weighted Bonferroni has higher dPoS than unweighted Bonferroni, with observed increases of up to 0.075. As expected, the difference in dPoS is symmetric about the line $\theta_1 = \theta_2$. The largest differences in dPoS occur in the region $\{0.5 < \theta_1, \theta_2 < 4\}$. When $\theta_1$ and $\theta_2$ are both small ($<0.5$), the dPoS is so low that weighting makes almost no difference. Conversely, when $\theta_1$ and $\theta_2$ are both large ($>4$) the dPoS is essentially equal to 1 when using the unweighted Bonferroni procedure, and so again weighting makes almost no difference. 

Starting with a fixed small value of $\theta_2 < 1$ and increasing the value of $\theta_1$, the difference in dPoS increases steadily from 0 to a maximum of approximately 0.07 when $\theta_2 \approx 2.5$, reflecting how the weight $\hat{w}_1$ smoothly increases from 0.5 towards 1. However, the difference in dPoS then starts to decrease despite the  weight $\hat{w}_1$ continuing to increase towards 1. This makes intuitive sense as when $\theta_1$ is very large (and $\theta_2$ is small), the weighted and unweighted Bonferroni procedures will both have a marginal PoS very close to 1 for $H_1$. Hence, there is a `sweet spot' around $\theta_2 \approx 2.5$ where the weighted Bonferroni approach makes the most difference in terms of dPoS.

Interestingly, the difference in dPoS is not zero when $\theta_1 = \theta_2$ even though on average the empirical weights $\hat{w}_1, \hat{w}_2$ are both equal to 0.5 (i.e.\ the unweighted case). For example, when $\theta_1 = \theta_2 = 1.7$ there is a difference in dPoS of 0.05. The reason for this is that the empirical weights $\hat{w}_1, \hat{w}_2$ take a range of values in $[0,1]$ (as shown in Figure~\ref{fig:2T_weight_histogram}), and despite this distribution being symmetric about 0.5, the implications of weights being less than $0.5$ and greater than $0.5$ are not symmetric. Focusing on $H_1$ (a similar argument holds for $H_2$), consider the case when $T_1^{(1)} < T_2^{(1)}$ and so $\hat{w}_1 <0.5$. The relative effect on the marginal PoS will be mitigated however, since if $T_1^{(1)}$ is small enough, then $H_1$ will fail to be rejected in trial~1 and hence will fail to be rejected overall regardless of whether a weighted or unweighted Bonferroni procedure is used in trial~2. In contrast, now consider the case when  $T_1^{(1)} > T_2^{(1)}$ and so $\hat{w}_1 > 0.5$. The relative effect on the marginal PoS is more substantial in this case, since if $T_1^{(1)}$ is large enough, then $H_1$ will be rejected in trial~1, $\hat{w}_1$ will be close to (or equal to) 1 and the power to reject $H_1$ in trial~2 (and hence the marginal PoS) will be noticeably higher when using the weighted compared with the unweighted Bonferroni procedure.

\begin{figure}[ht!]
\includegraphics[width = \linewidth]{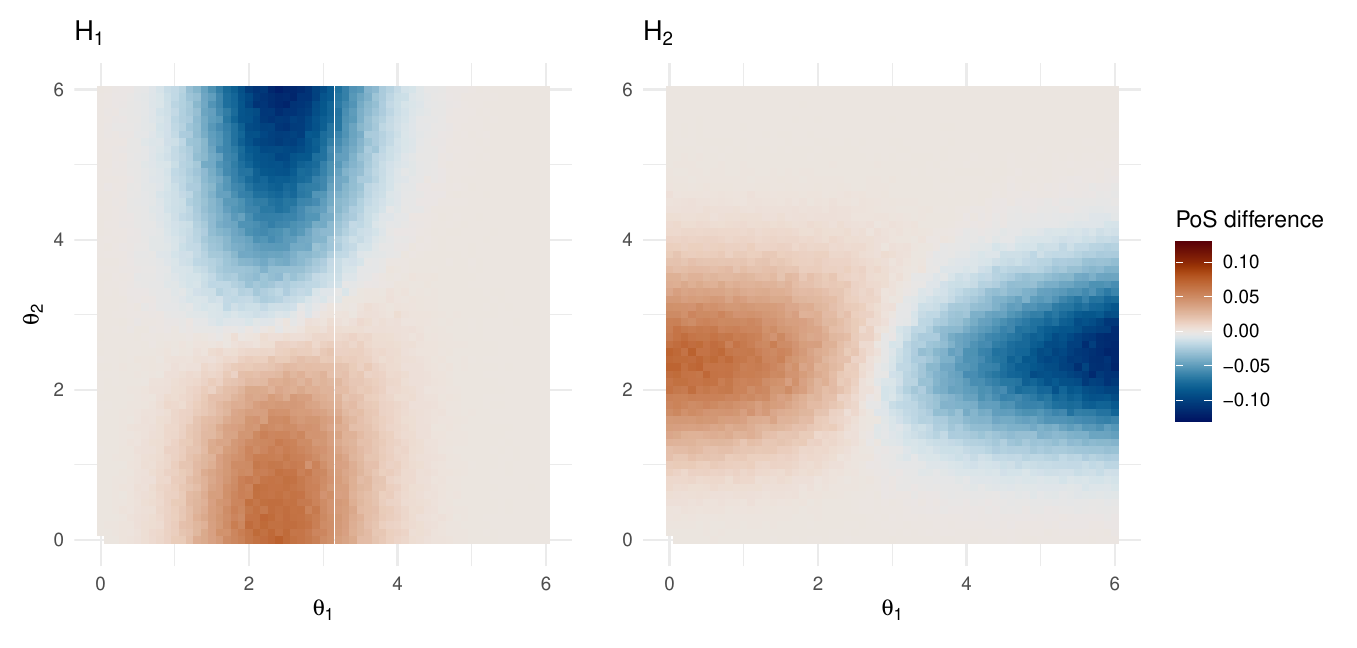} 
\centering
\caption{Heatmap showing difference in the marginal Probability of Success (PoS) for $H_1$ and $H_2$ for $\bm{\theta} = (\theta_1, \theta_2)$ where $0 \leq \theta_1, \theta_2, \leq 6$. Results are from $10^4$ simulations for each pair of values of $(\theta_1, \theta_2)$. Positive values indicate where the PoS of the weighted Bonferroni approach is higher than the PoS of the unweighted Bonferroni approach. \label{fig:mPoS_heatmap}}
\end{figure}

While there is a uniform improvement in the disjunctive PoS, the picture is not so clear cut when it comes to the \textit{marginal} PoS for $H_1$ and $H_2$, as seen in the heatmap given in Figure~\ref{fig:mPoS_heatmap}.

In the region $\{0 < \theta_1, \theta_2 < 2.5\}$ the weighted Bonferroni procedure has a higher PoS than the unweighted version for both $H_1$ and $H_2$, with essentially no difference in the PoS in the regions $\{\theta_1, \theta_2 > 4.5\}$, $\{\theta_1 < 0.5, \theta_2 > 2.5\}$ and $\{\theta_1 > 2.5, \theta_2 < 0.5\}$. However, for the region $\{0.5 < \theta_1 < 4.5, \theta_2 > 2.5\}$ the PoS for $H_1$ is lower when using weighted Bonferroni compared with unweighted Bonferroni, and similarly for the region $\{0.5 < \theta_2 < 4.5, \theta_1 > 2.5\}$ the PoS for $H_2$ is lower. Looking at the heatmap of empirical mean weights $\hat{w}_1$ in Figure~\ref{fig:2T_weight_heatmap}, the region where the PoS for $H_1$ is lower corresponds to where $\theta_1$ is in the `interesting' zone (i.e.\ not too small or too large) but $\theta_2$ is larger and so $\hat{w}_1 < 0.5$ (a similar argument holds for the region where the PoS for $H_2$ is lower). This is the price that has to be paid to maximise the disjunctive PoS, although reassuringly the more substantial losses in marginal PoS only occur for very large values of $\theta_1$ or $\theta_2$ which coincide with very high values of disjunctive PoS anyway and (depending on the disease area) may be rather unlikely to be seen in practice.

\begin{figure}[ht!]
\includegraphics[width = 0.7\linewidth]{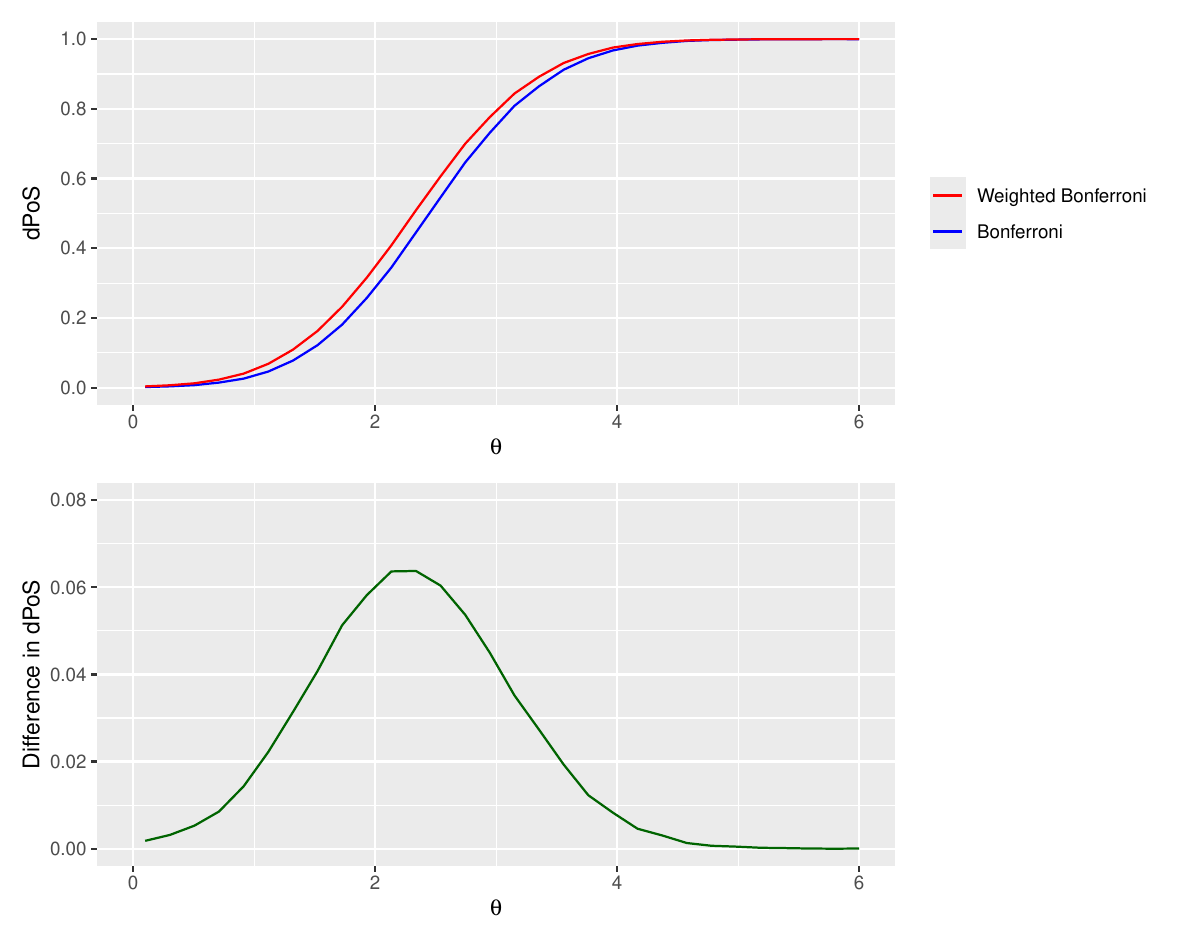} 
\centering
\caption{Disjunctive Probability of Success (dPoS) for $\bm{\theta} = (\theta/2, \theta)$. Results are from $10^5$ simulation replicates.\label{fig:dPoS_2T_sim2}}
\end{figure}

We now take a closer look at the setting where $\bm{\theta} = (\theta/2, \theta)$ with $0 < \theta < 6$. Figure~\ref{fig:dPoS_2T_sim2} shows the disjunctive PoS for the weighted and unweighted Bonferroni procedures (and their difference), averaged across $10^5$ simulation replicates. As already observed from Figure~\ref{fig:2T_weight_heatmap}, the difference in the disjunctive PoS increases as $\theta$ increases, reaching a maximum of about 0.06 when $\theta \approx 2.3$, which then starts to decrease. Looking at the upper plot we can see that the weighted Bonferroni procedure has a noticeable gain in PoS in regions that matter, i.e. where the PoS is relatively high but not quite reaching conventional levels of 80\%.

\begin{figure}[ht!]
\includegraphics[width = \linewidth]{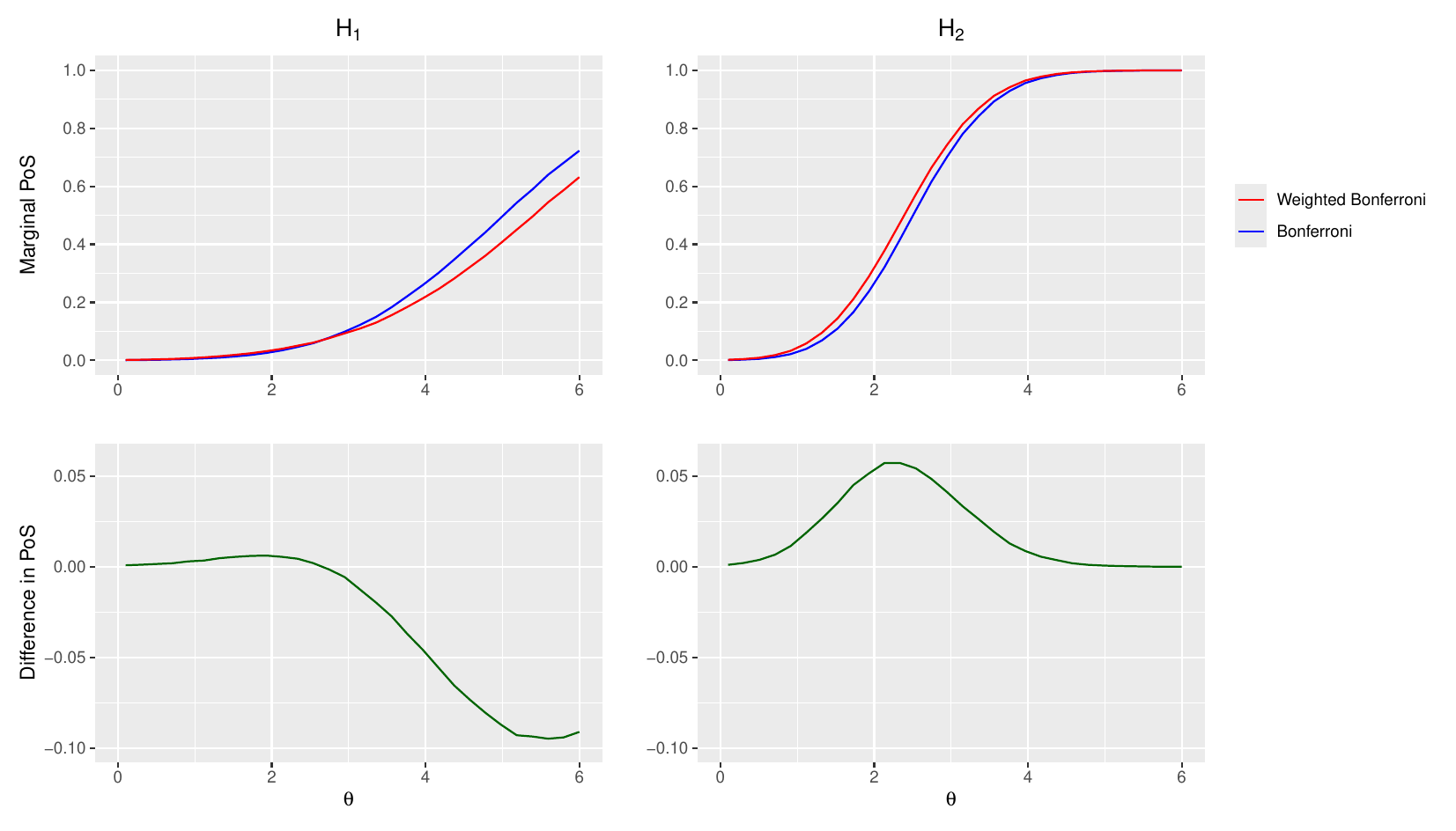}
\centering
\caption{Marginal Probability of Success (PoS) for $H_1$ and $H_2$ for $\bm{\theta} = (\theta/2, \theta)$. Results are from $10^5$ simulation replicates. \label{fig:mPoS_2T_sim2}}
\end{figure}

Figure~\ref{fig:mPoS_2T_sim2} shows the \textit{marginal} PoS for $H_1$ and $H_2$ in the same setting. We can see that the gains in disjunctive PoS are almost all driven by the gains in the marginal PoS for $H_2$, with the two plots closely matching. In contrast, the marginal PoS for $H_1$ is only very slightly higher for the weighted Bonferroni procedure for $0 < \theta < 2.5$, while for $\theta > 2.5$ the weighted Bonferroni procedure has an increasingly lower PoS than the unweighted Bonferroni procedure, as already seen in Figure~\ref{fig:mPoS_heatmap}.

Table~\ref{tab:rej_sim2b} shows the detailed results for rejection and non-rejection for the two hypothesis for $\theta=3$ based on 1000 simulations. The vast majority of the time the weighted and unweighted Bonferroni procedures make the same overall rejections (973/1000 and 949/1000 for $H_1$ and $H_2$, respectively). There were 51/1000 cases of additional overall rejections of $H_2$ when using the weighted Bonferroni procedure. In contrast, there were 23/1000 additional overall rejections of $H_1$ when using the unweighted Bonferroni procedure.

%
%
%

\vspace{12pt}

\begin{table}[ht!]
\scriptsize
\centering
\begin{minipage}{0.49\textwidth} 
\centering
\textbf{H1} \\[12pt] 
\begin{tabular}{c c | c c}
\multicolumn{2}{c|}{\textbf{Test}} & \multicolumn{2}{c}{\textbf{Weighted}} \\
 & & Rejection & No rejection \\
\hline
\multirow{2}{*}{\textbf{Unweighted}} & Rejection & 101  & 4 \\
& No rejection & 23 & 872
\end{tabular}
\end{minipage}
\hfill 
\begin{minipage}{0.49\textwidth} 
\centering
\textbf{H2} \\[12pt] 
\begin{tabular}{c c | c c}
\multicolumn{2}{c|}{\textbf{Test}} & \multicolumn{2}{c}{\textbf{Weighted}} \\
 & & Rejection & No rejection \\
\hline
\multirow{2}{*}{\textbf{Unweighted}} & Rejection & 715  & 0 \\
& No rejection & 51 & 234
\end{tabular}
\end{minipage}

\vspace{8pt} 

\caption{\small Overall rejections of $H_1$ and $H_2$ when using the weighted and unweighted Bonferroni procedures for $\bm{\theta} = (3/2, 3)$. Results are from $1000$ simulation replicates. \label{tab:rej_sim2b}}
\end{table}

Despite the guaranteed gains in disjunctive PoS, as we have seen there can be losses in terms of marginal PoS. The scenario where $\bm{\theta} = (\theta/2, \theta)$ gives large losses in the marginal PoS (for $H_1$) when $\theta \approx 5$, but this is not the case in other scenarios as we briefly describe below.


One special case is the scenario where $\bm{\theta} = (0, \theta)$, where the marginal PoS is the same as the disjunctive PoS, since only $H_2$ is non-null. Figure~\ref{fig:PoS_2T_sim1} in the Appendix shows the marginal PoS for $H_2$ using the unweighted and weighted Bonferroni procedures. The weighted Bonferroni procedure always has a higher marginal PoS than the unweighted procedure, with a maximum increase of almost 0.07 when $\theta \approx 2.3$. 
Taking $\theta = 3$, we can also perform a comparison of the number of rejections of $H_1$ and $H_2$, with full results are shown in Table~\ref{tab:rej_sim2a} in the Appendix, taken from $1000$ simulation replicates.
%
In 50/1000 cases the weighted Bonferroni procedure rejected $H_2$ overall whereas the unweighted Bonferroni procedure did not. There were no cases were the unweighted Bonferroni procedure rejected $H_2$ overall and the weighted Bonferroni procedure did not, illustrating the gain in PoS.


Finally we consider the scenario with equal treatment means, i.e.\ where $\bm{\theta} = (\theta, \theta)$ with $\theta > 0$. Figure~\ref{fig:dPoS_2T_sim3} in the Appendix shows the disjunctive PoS and the difference between the weighted and unweighted Bonferroni procedures, while Figure~\ref{fig:mPoS_2T_sim3} shows the results for the marginal PoS for $H_1$ and $H_2$ (which are identical up to simulation error). The gains in the marginal PoS are when $\theta < 2.5$, with the difference in PoS again becoming negative for $\theta > 2.5$. However, this time the maximal loss is low, at less than 0.005. Taking $\theta = 3$, Table~\ref{tab:rej_sim2c} shows the number of rejections of $H_1$ and $H_2$ from $1000$ simulation replicates.  
There are substantially more cases where the weighted Bonferroni procedure rejects and the unweighted Bonferroni procedure does not reject compared with the other way round, illustrating the minimal loss in marginal PoS.



\subsubsection{$m>2$ treatments}
\label{subsec:multiple_treatments}

\begin{figure}[ht!]
\includegraphics[width = 0.7\linewidth]{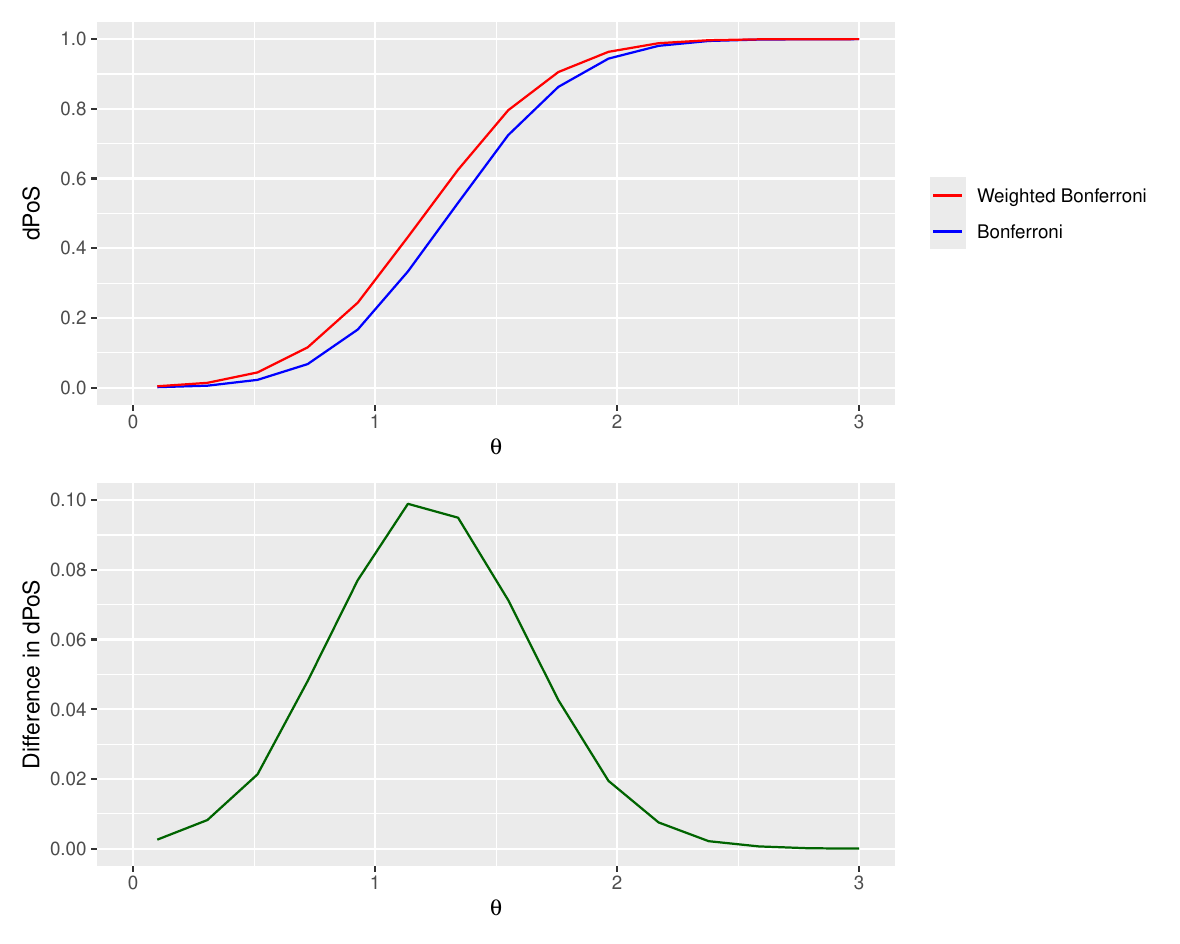} 
\centering
\caption{Disjunctive Probability of Success (dPoS) for $\bm{\theta} = (\theta/2, \theta, 2\theta)$. Results are from $10^5$ simulation replicates. Positive values of the difference in dPoS indicate where the dPoS of the weighted Bonferroni approach is higher than the unweighted Bonferroni approach.\label{fig:dPoS_3T_sim2}}
\end{figure}
\begin{figure}[ht!]
\includegraphics[width = \linewidth]{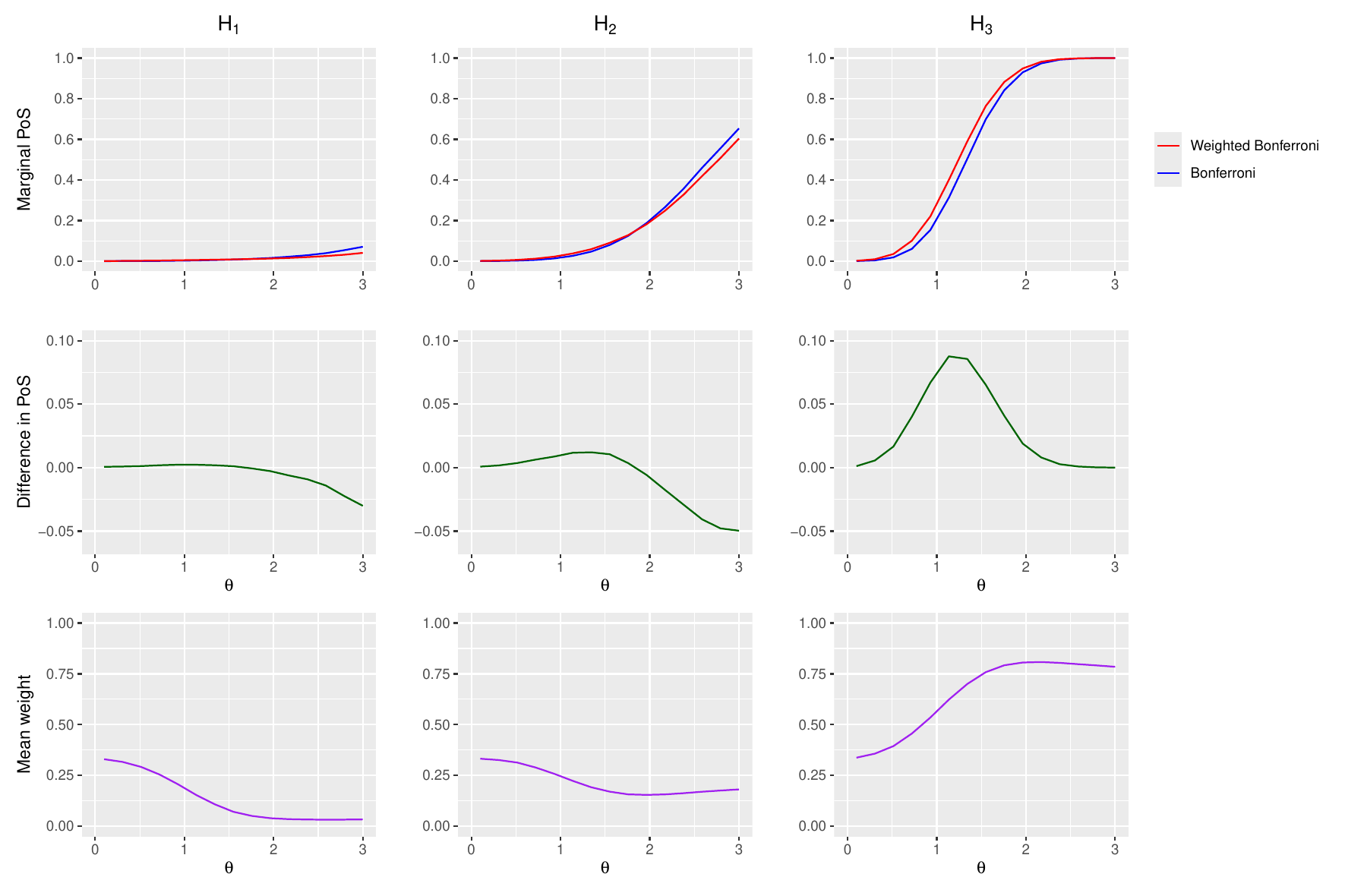} 
\centering
\caption{Marginal Probability of Success (PoS) and mean empirical weights for $H_1, H_2,  H_3$ for $\bm{\theta} = (\theta/2, \theta, 2\theta)$. Results are from $10^5$ simulation replicates. Positive values of the difference in PoS indicate where the PoS of the weighted Bonferroni approach is higher than the unweighted Bonferroni approach.\label{fig:mPoS_3T_sim2}}
\end{figure}

Moving to the setting with three treatments, we focus on the scenario $\bm{\theta} = (\theta/2, \theta, 2\theta)$ with $0 < \theta < 3$ (we only go up to $\theta = 3$ since the disjunctive PoS is already equal to 1).  Figure~\ref{fig:dPoS_3T_sim2} shows the disjunctive PoS, while Figure~\ref{fig:mPoS_3T_sim2} shows the marginal PoS for $H_1, H_2$ and $H_3$ as well as the mean (empirical) weights $\hat{w}_1, \hat{w}_2, \hat{w}_3$. 

In terms of the disjunctive PoS, the weighted Bonferroni procedure always has higher values than the unweighted Bonferroni procedure. There is a maximum gain of about 0.1 when $\theta \approx 1.1$, with noticeable gains in the region where the disjunctive PoS is quite high but not quite at the conventional 0.8 or 0.9 level. This gain in disjunctive PoS is almost entirely driven by the large gains possible in marginal PoS for $H_3$. In contrast, for the marginal PoS for $H_2$ the weighted Bonferroni procedure performs worse than unweighted Bonferroni when $\theta > 1.8$, with a difference up to $-0.05$ for $\theta = 3$, reflecting how the weights $\hat{w}_2$ remain less than 1/3. Similarly, the difference in marginal PoS for $H_1$ is also negative when $\theta > 1.8$, with the weights becoming close to zero. 


Like for the two treatment setting, this loss in marginal PoS (despite the guaranteed gain in the disjunctive PoS) is not seen in other scenarios. For example, consider the scenario where $\bm{\theta} = (0, 0, \theta)$. The marginal PoS for $H_3$ (which is the same as the disjunctive PoS) is always greater when using the weighted Bonferroni procedure compared with the unweighted procedure, with a maximal gain in PoS of 0.101. 


%

\begin{figure}[ht!]
\includegraphics[width = 0.8\linewidth]{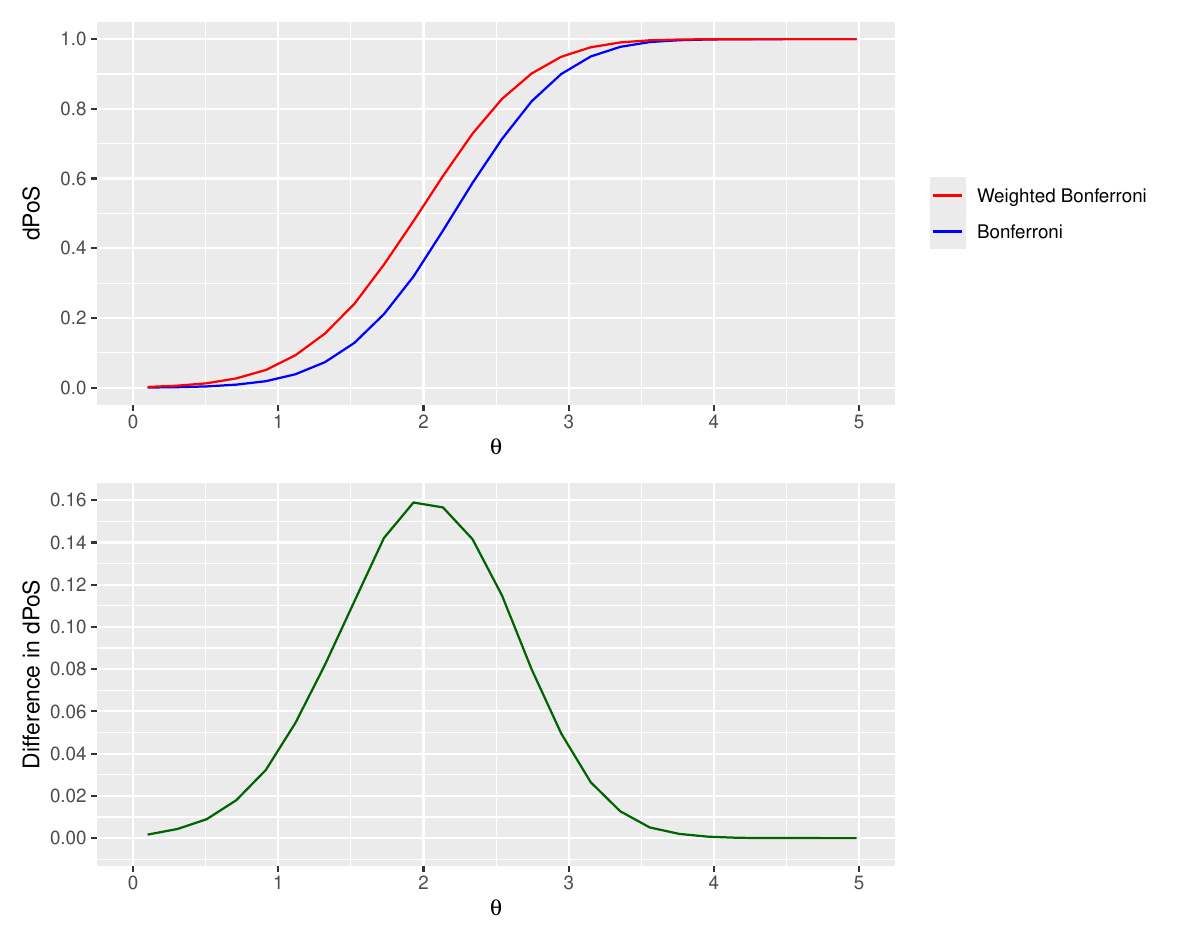} 
\centering
\caption{Disjunctive Probability of Success (dPoS) for $\bm{\theta} = (0, 0, \theta, \theta, \theta)$. Results are from $10^5$ simulation replicates.  \label{fig:dPoS_5T_sim2}}
\end{figure}

Finally, we consider the setting with five treatments. 
We focus on the scenario,  $\bm{\theta} = (0, 0, \theta, \theta, \theta)$, where $0 < \theta < 5$. Figure~\ref{fig:dPoS_5T_sim2} shows the disjunctive PoS, where large gains when using the weighted Bonferroni procedure are seen (up to 0.16). This includes interesting regions of the parameter space when using the weighted Bonferroni procedure. For example, when $\theta \approx 2.5$ then the PoS for the weighted Bonferroni procedure is approximately 0.80, compared with around 0.70 when using unweighted Bonferroni. 

\begin{figure}[ht!]
\includegraphics[width = 0.8\linewidth]{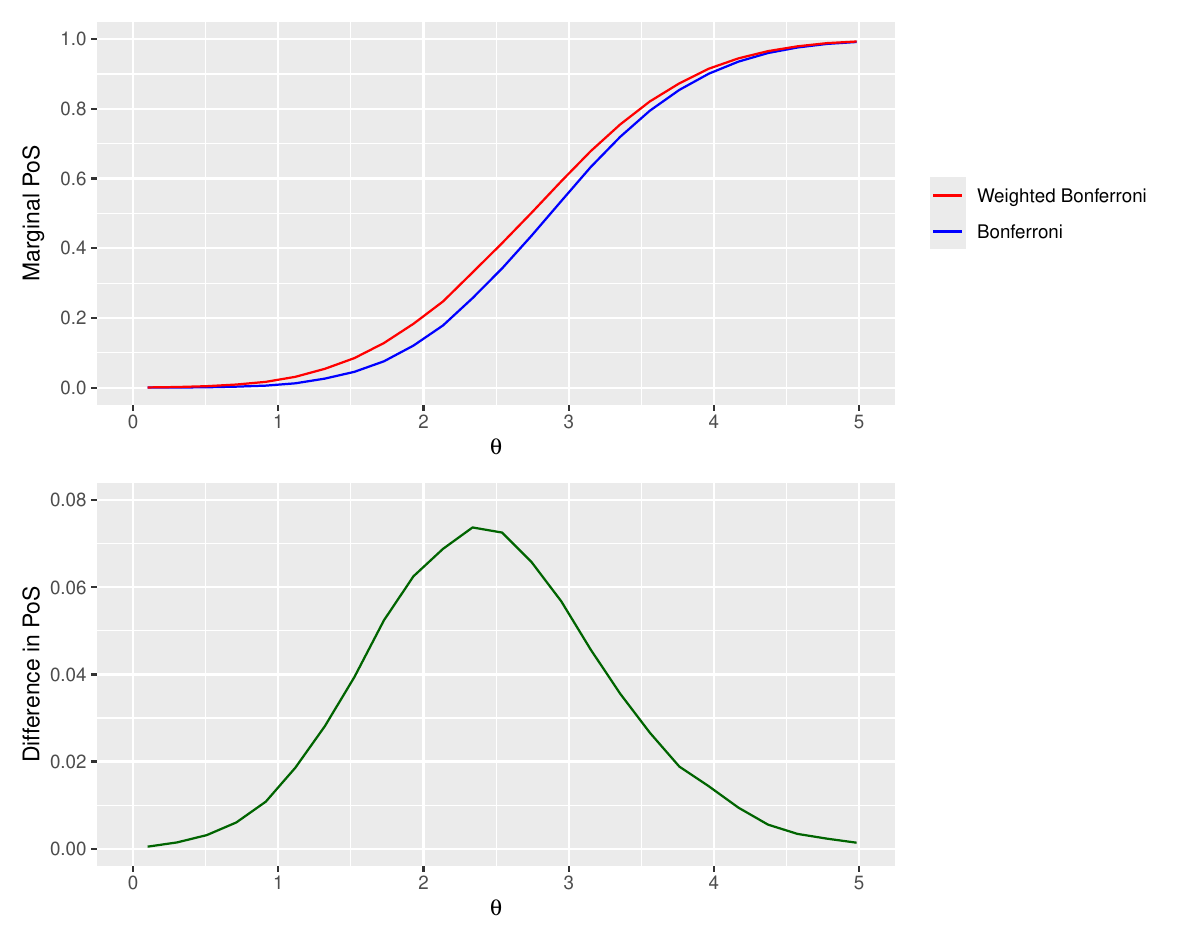} 
\centering
\caption{Marginal Probability of Success (PoS) for $H_5$ for $\bm{\theta} = (0, 0, \theta, \theta, \theta)$. Results are from $10^5$ simulation replicates.  \label{fig:mPoS_5T_sim2}}
\end{figure}

Figure~\ref{fig:mPoS_5T_sim2} shows the marginal PoS for $H_5$, which will be the same plot as the marginal PoS for any of the non-nulls $H_3, H_4, H_5$ (apart from simulation error). There are substantial gains in the marginal PoS for any of the non-nulls (up to 0.074), with weighted Bonferroni always having a higher marginal PoS than unweighted Bonferroni.

Even larger gains in the marginal PoS (of up to 0.134) are seen in the scenario where $\bm{\theta} = (0, 0, 0, 0, \theta)$, as seen in Figure~\ref{fig:mPoS_5T_sim1} in the Appendix which shows the marginal PoS for $H_5$ (which is the same as the disjunctive PoS in this case). Finally, a more complex scenario is given by $\bm{\theta} = (\theta_1, \theta_2, \theta_3, \theta_4,\theta)$, with $\theta_i \sim U[0, \theta]$ independently for $i = 1, 2, 3, 4$ and $0 < \theta< 5$. Figure~\ref{fig:dPoS_5T_sim3} in the Appendix shows the disjunctive PoS, with the figure for the marginal PoS for $H_5$ given in~Figure~\ref{fig:mPoS_5T_sim3} in the Appendix. For the disjunctive PoS, the results are similar as for the previous scenarios with five treatments, with a maximum gain of 0.149. In terms of the marginal PoS, at least for the largest treatment effect (corresponding to $H_5$) we again see that there is a substantial gain in PoS, up to 0.096. 

\subsection{Robustness}
\label{subsec:robustness}

The assumption that experiments~1 and~2 have exactly the same underlying true parameters $\theta_1, \ldots, \theta_m$ is  a strong one, so in the following simulation studies we also consider what happens when this assumption is violated. That is, the setting where $T_i^{(1)} \sim N(\theta_i, 1)$ but $T_i^{(2)} \sim N(\theta_i', 1)$ where $\theta_i' \neq \theta_i$.  Such discrepancies could occur for a number of reasons, including due to systematic differences in prognostic variables between the populations assessed in experiments~1 and~2, and/or some form of temporal trend that causes the results in the later experiment~2 to be systematically different from the earlier experiment~1. In the context of replication studies, another potential feature is \textit{publication bias}, so that the results from experiment~1 are in fact biased high (i.e., to give a `positive' result that is published, when in fact the true effect is null). 

\begin{figure}[ht!]
\includegraphics[width = 0.8\linewidth]{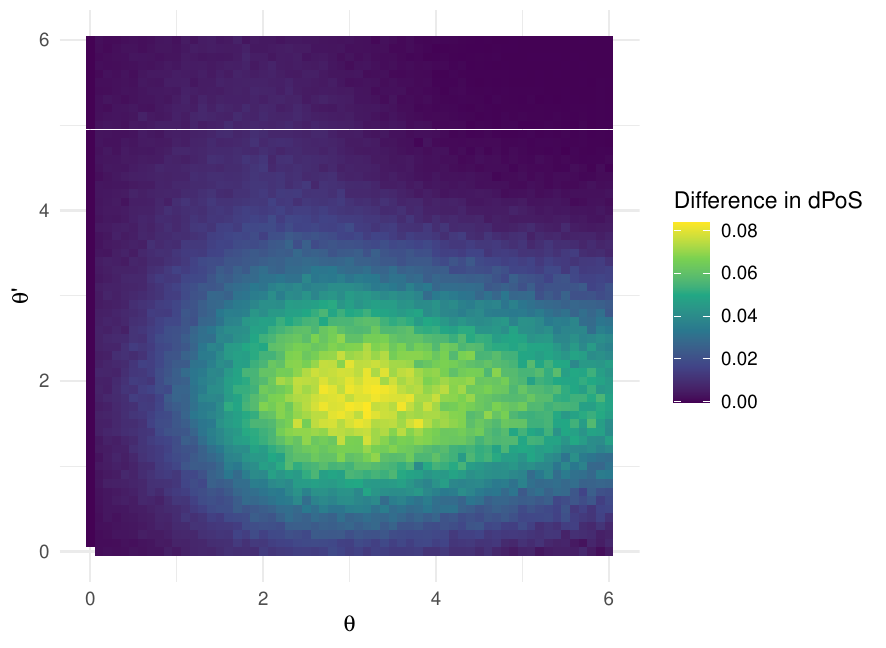} 
\centering
\caption{Heatmap showing difference in the disjunctive Probability of Success (dPoS) when trial~1 has treatment means given by $\bm{\theta} = (\theta/2, \theta)$ while trial~2 has treatment means given by $\bm{\theta'} = (\theta'/2, \theta')$, where $0 \leq \theta, \theta' \leq 3$. Results are from $10^4$ simulations for each pair of values of $(\theta, \theta')$. Positive values indicate where the PoS of the weighted Bonferroni approach is higher than the PoS of the unweighted Bonferroni approach. \label{fig:dPoS_2T_heatmap_robust_b}}
\end{figure}

To assess robustness, we focus on the two treatment setting for simplicity and assume trial~1 has treatment means $\bm{\theta} = (\theta/2, \theta)$ while trial~2 has treatment means $\bm{\theta'} = (\theta' /2, \theta')$, where $0 \leq \theta, \theta' \leq 6$. Figure~\ref{fig:dPoS_2T_heatmap_robust_b} shows a heatmap of the difference in the disjunctive PoS between the weighted and unweighted Bonferroni procedures. Despite the sometimes very large differences between $\bm{\theta}$ and $\bm{\theta'}$, the weighted Bonferroni procedure still always has equal or higher PoS compared with the unweighted Bonferroni procedure. 
Interestingly, this misspecified model with $\theta' \neq \theta$ can lead to \textit{larger} gains in PoS compared with when $\theta' = \theta$ for some regions of the parameter space.

\begin{figure}[ht!]
\includegraphics[width = \linewidth]{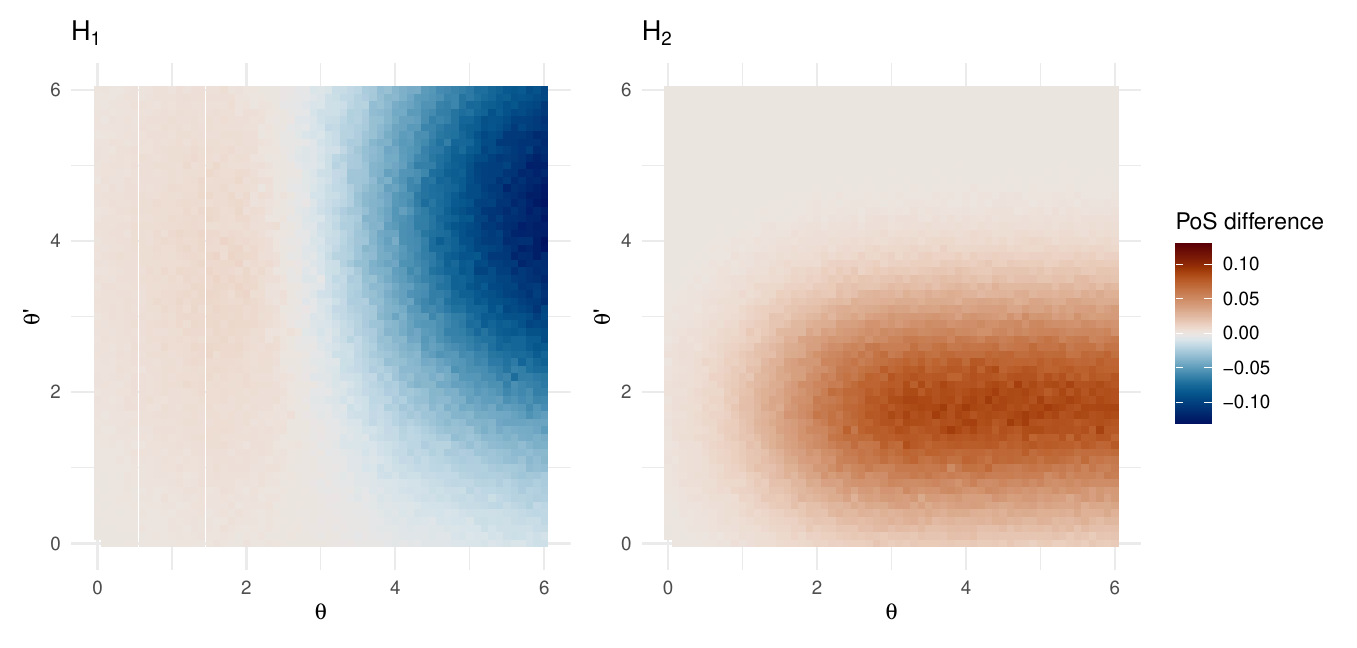} 
\centering
\caption{Heatmap showing difference in the marginal Probability of Success (PoS) for $H_1$ and $H_2$ when trial~1 has treatment means given by $\bm{\theta} = (\theta/2, \theta)$ while trial~2 has treatment means given by $\bm{\theta'} = (\theta'/2, \theta')$, where $0 \leq \theta, \theta' \leq 6$. Results are from $10^4$ simulations for each pair of values of $(\theta, \theta')$. Positive values indicate where the PoS of the weighted Bonferroni approach is higher than the PoS of the unweighted Bonferroni approach. \label{fig:mPoS_2T_heatmap_robust_b}}
\end{figure}

In terms of the difference in the marginal PoS, as seen in Figure~\ref{fig:mPoS_2T_heatmap_robust_b}, for $H_2$ the misspecified means still do not lead to a loss in marginal PoS compared with the unweighted Bonferroni procedure. For the region $\{\theta > 1, \ < \theta' < 3\}$ we in fact see an increase in the PoS difference compared to when $\theta = \theta'$. As for the marginal PoS for $H_1$, when $\theta<2.5$ there is minimal difference between the PoS. However for $\theta > 2.5$ then regardless of the value of $\theta'$, there is a loss in PoS, although this is worst in the region $\{\theta > 4, \theta'> 2\}$.

\begin{figure}[ht!]
\includegraphics[width = 0.65\linewidth]{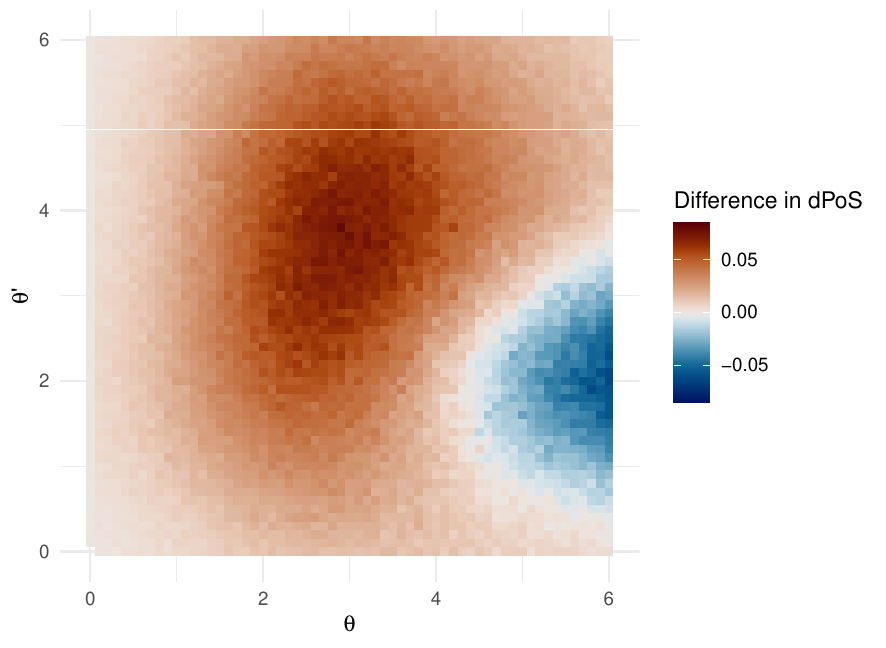} 
\centering
\caption{Heatmap showing difference in the disjunctive Probability of Success (dPoS) when trial~1 has treatment means given by $\bm{\theta} = (\theta/2, \theta)$ but trial~2 has treatment means given by $\bm{\theta'} = (\theta', \theta'/2)$, where $0 \leq \theta, \theta' \leq 6$. Results are from $10^4$ simulations for each pair of values of $(\theta, \theta')$. Positive values indicate where the PoS of the weighted Bonferroni approach is higher than the PoS of the unweighted Bonferroni approach. \label{fig:dPoS_2T_heatmap_robust}}
\end{figure}

To find a scenario where the weighted Bonferroni procedure has a lower disjunctive PoS than the unweighted Bonferroni procedure, we need to be more extreme in the difference in the means between the two trials. To this end, we again assume trial~1 has treatment means $\bm{\theta} = (\theta/2, \theta)$, but this time assume trial~2 has treatment means $\bm{\theta'} = (\theta', \theta'/2)$ where $0 \leq \theta, \theta' \leq 6$. Figure~\ref{fig:dPoS_2T_heatmap_robust} shows the resulting heatmap of the difference in disjunctive PoS. This time, in the region centred around $\theta' = 2$ and $\theta > 4$ there is a decrease in disjunctive PoS of up to 0.063, driven by a relatively large drop in the marginal PoS of $H_1$ due to the badly misspecified weights. 
Arguably though the magnitude of the discrepancy between the two trials where this happens is large enough to be unrealistic in many trial scenarios.

Further results for robustness in the two treatment setting can be found in the Appendix. Figure~\ref{fig:dPoS_2T_heatmap_robust_a} shows a heatmap of the difference in the disjunctive PoS (i.e.\ the PoS for $H_2$) between the weighted and unweighted Bonferroni procedures when trial~1 has treatment means $\bm{\theta} = (0, \theta)$ while trial~2 has treatment means $\bm{\theta'} = (0, \theta')$. Again we see that the disjunctive PoS of the weighted Bonferroni procedure is always equal or higher than that of the unweighted Bonferroni procedure. Finally, we consider the setting where trial~1 has treatment means $\bm{\theta} = (\theta, \theta)$ while trial~2 has treatment means $\bm{\theta'} = (\theta', \theta')$. The resulting heatmap of the difference in disjunctive PoS is given in Figure~\ref{fig:dPoS_2T_heatmap_robust_c}. Throughout the whole parameter space the dPoS of the weighted Bonferroni procedure is higher than unweighted Bonferroni, except for the region $\{ \theta > 3, \theta' < 3\}$ where the difference can go very slightly negative (at most $-0.005$).


\section{Discussion}
\label{sec:discuss}


The use of the proposed weighted Bonferroni procedure can lead to a substantial gain in the disjunctive PoS (compared with using the usual unweighted Bonferroni procedure) in the context of the `two-trial' rule for confirmatory clinical trials. Importantly, such gains can be realised in `interesting' regions of the parameter space where the disjunctive PoS is relative high when using the unweighted Bonferroni procedure, but not quite reaching conventional levels of 80\% or 90\%. The magnitude of these gains also increases with the number of tested hypotheses~$m$ (i.e.\ treatment options), since there is more scope to deviate from the usual equal weighting of $(1/m, \ldots, 1/m)$. For example, the maximum gain in the disjunctive PoS increases from 0.069 to 0.101 to 0.134 for $\bm{\theta} = (0, \theta)$,  $\bm{\theta} = (0, 0, \theta)$ and $\bm{\theta} = (0, 0, 0, 0, \theta)$, respectively.

When the true treatment means in trials~1 and~2 are the same, weighted Bonferroni has uniformly higher disjunctive PoS than unweighted Bonferroni, as would be expected from theory. However, even when the true treatment means in trials~1 and~2 differ, the weighted Bonferroni procedure is surprisingly robust in terms of still having higher disjunctive PoS, at least in the two treatment setting. We would expect this robustness to also be seen with higher number of treatments, although to a lesser extent given the greater scope for the relative orderings of treatment effect to (dramatically) change between trial~1 and~2. 

As seen in the simulation results, there is often a trade-off between  maximising the disjunctive PoS (and hence the marginal PoS for treatments with the highest trial~1 means) and loss in marginal PoS for treatments with more `moderate' treatment effects in trial~1. This is somewhat inevitable given that the optimisation criteria used is the disjunctive power of trial~2, meaning that the treatments with the highest trial~1 means are given the highest weights and hence treatments with lower trial~1 means must be given lower weights. If the interest is in the marginal PoS of more/all treatments in the trial, then other optimisation criteria may be more appropriate to consider and use for the weighting of the procedure. As briefly mentioned in the Introduction, such optimisation criteria include the mean power or conjunctive power of trial~2. However, these alternative optimality criteria will also have their own potential trade-offs and the choice depends on the trial context and aims.

As a simplifying assumption, in this paper we have considered the setting where the test statistics (or equivalently, the $p$-values) in trial~2 are independent. In practice, depending on the trial context, these test statistics may be correlated (for example, if each hypothesis corresponds to a different endpoint measured on the same group of patients). It is important to note that even if the test statistics have some arbitrary correlation structure (even negative correlations), the proposed weighted Bonferroni procedure will still guarantee FWER control for trial~2 by definition. It is in the disjunctive power of trial~2 that ignoring the correlation structure will make an impact. Conceivably, using the weighted Bonferroni procedure assuming independence could result in a lower disjunctive power in trial~2 (and hence lower disjunctive PoS) then using the unweighted Bonferroni procedure. However, the results in~\cite{Xi2024} suggest that it is only with very high correlation values (i.e.\ $>0.7$) that the optimal weights change (at least in the setting with equal correlations and equal treatment means). In any case, if (highly) correlated test statistics are a concern, then the methods proposed in~\cite{Xi2024} for finding optimal weights for the disjunctive power accounting for correlation could be used.

Finally, we have focused on weighted Bonferroni procedures in this paper. There are a number of reasons for this, including simplicity, ease of interpretability in terms of adjusted $p$-values and confidence intervals, and robustness to deviations from independence in terms of still guaranteeing FWER control. However, an alternative would be to consider weighted Holm procedures (also known as Bonferroni-Holm), which is a stepdown procedure that also guarantees FWER control for arbitrary correlation between the test statistics. This would require a reformulation of the optimisation problem to take into account the more complex procedure used in trial~2.
Similarly, if it is known that the test statistics are positively correlated then one could consider using a weighted Hochberg procedure. Finally, if the full correlation structure was known in terms of a multivariate normal distribution (which would be the case if we are comparing multiple independent normally-distributed treatment arms against a common control, for example) the a weighted Dunnett test could be considered instead.


\bibliography{optimal_weight}

\section*{Appendix}
\appendix
\counterwithin{figure}{section}

\section{Proof of form of optimal weights}
\label{Asec:proof}

To solve the optimisation problem given in~\eqref{eq:optim_problem}, we form the Lagrangian $\mathcal{L}$ as follows: \[
\mathcal{L} = 1 - \prod_{i = 1}^m \left[ \Phi \! \left( \bar{\Phi}^{-1} (w_i \alpha ) - \theta_i \right) \! \mathbbm{1} (\theta_i > 0)\right] - \lambda \! \left( \! m - \sum_{i=1}^m w_i \mathbbm{1} (\theta_i > 0) \! \right).
\]
Let $\phi$ denote the standard normal PDF (probability density function). Taking the (partial) derivative of $\mathcal{L}$ with respect to $w_j$, where $\theta_j > 0$ (we simply set $w_j = 0$ otherwise) and equating this to zero gives 
\begin{align*}
& \frac{\partial \mathcal{L}}{\partial w_j}  = \lambda - \prod_{i \in \mathcal{A}, \, i \neq j} \alpha \Phi \! \left( \bar{\Phi}^{-1} (w_i \alpha ) - \theta_i \right) \! \mathbbm{1} (\theta_i> 0) \left[ \frac{\phi \! \left( \bar{\Phi}^{-1} (w_j \alpha ) - \theta_j  \right)}{\phi \! \left(  \bar{\Phi}^{-1} (w_j \alpha) \right) } \right] = 0 \\
& \implies  \frac{\phi \! \left( \bar{\Phi}^{-1} (w_j \alpha ) - \theta_j  \right)}{\phi \! \left(  \bar{\Phi}^{-1} (w_j  \alpha) \right) } = \frac{\lambda}{\alpha} \left[ \prod_{i \in \mathcal{A}, \, i \neq j} \! \mathbbm{1} (\theta_i> 0) \, \Phi \! \left( \bar{\Phi}^{-1} (w_i \alpha) - \theta_i \right) \right]^{-1} \\
& \implies  \exp \! \left(-\frac{\theta_j^2}{2} + \theta_j \bar{\Phi}^{-1} (w_j \alpha) \right) =   \frac{\lambda}{\alpha} \left[ \prod_{i \in \mathcal{A}, \, i \neq j} \! \mathbbm{1} (\theta_i> 0) \, \Phi \! \left( \bar{\Phi}^{-1} (w_i \alpha ) - \theta_i \right) \right]^{-1} \\
& \implies \bar{\Phi}^{-1} (w_j \alpha) = \frac{\theta_j}{2} + \frac{1}{\theta_j} \log \! \left[ \frac{ \lambda}{\alpha}\left( \prod_{i \in \mathcal{A}, \, i \neq j} \!\mathbbm{1} (\theta_i> 0) \, \Phi \! \left( \bar{\Phi}^{-1} (w_i \alpha ) - \theta_i \right) \right)^{-1} \; \right] \\
& \implies w_j =  \frac{1}{\alpha} \, \bar{\Phi} \! \! \left( \frac{\theta_j}{2} + \frac{1}{\theta_j} \! \left[ \log(c) - \! \!\sum_{i \in \mathcal{A} , \, i \neq j} \! \!\mathbbm{1} (\theta_i> 0) \log \! \left( \Phi (\bar{\Phi}^{-1} (w_i \alpha ) - \theta_i ) \right) \right] \right)
\end{align*}
where $c = \frac{\lambda}{\alpha}$.

\section{Comparison of computational methods}
\label{Asec:computational_methods}

In this Section, we compare the following numerical methods for finding optimal weights:
\begin{enumerate}
\item nleq = use of non-linear simultaneous equations solver (\texttt{nleqslv} package in R), as proposed and used in this paper.
\item nlopt = use of non-linear optimisation algorithm with constraints (\texttt{nloptr} package in R) and a single starting point, as proposed by~\cite{Xi2024}
\item nlopt multi = use of non-linear optimisation algorithm with constraints (\texttt{nloptr} package in R) and multiple starting points, as proposed by~\cite{Xi2024}
\item grid search
\end{enumerate}

We use the above numerical methods to find optimal weights for~$m$ treatments, $m \in \{2,3,4,5\}$), where the true treatment means are given by $\bm{\theta} = (\theta_1, \ldots, \theta_m)$ with $\theta_i \sim U[0, b]$ independently for $i = 1, \ldots, m$. Here the upper bound $b = 6$ for $m = 2$ and $b = 3$ otherwise. We generate 1000 such vectors of treatment means for each value of~$m$.

Table~\ref{tab:timings} shows the time taken for each method to calculate the optimal weights. The nlopt method is the fastest, but as we shall see comes at the cost of sub-optimal weights. The nleq method is about two times slower than nlopt. Finally, the nlopt multi method is substantially slower than either nleq or nlopt, particularly when $m > 2$, which is to be expected given that $2^m - 1$ starting points need to be considered for this method. 

\begin{table}[ht!]

\begin{tabular}{c|cccc}

\textbf{Method} & \textbf{Two treatments} & \textbf{Three treatments} & \textbf{Four treatments} & \textbf{Five treatments} \\ 
\hline 
\textbf{nleq} & 2.2 & 4.7 & 8.9 & 15.8 \\ 

\textbf{nlopt} & 1.5 & 2.5 & 4.0 & 6.3\\ 

\textbf{nlopt multi} & 4.6 & 28.4 & 119.9& 472.5\\ 

\textbf{grid search} & 1.3 & 338.8 & -- & --

\end{tabular} 

\caption{Timings in seconds to calculate optimal weights for 1000 sets of uniform random treatment effect vectors. For grid search, the grid size was set equal to 0.005.\label{tab:timings}}

\end{table}

Table~\ref{tab:timings} compares the optimal weights and resulting disjunctive power found using nleq, nlopt and nlopt multi with those found using grid search for $m = 2,3$. The nleq method always finds the same optimal weights as those found using grid search (rounding to the grid size) and always has the greater or equal disjunctive power. In contrast, nlopt and nlopt multi find suboptimal weights (compared with those found using grid search) for a noticeable proportion of treatment effect vectors, particularly for $m = 2$.

The absolute magnitude of the power differences is small, however. The maximum power loss compared to using grid search for $m=2$ was $0.0039\%$ and $0.0011\%$ (in absolute terms) for nlopt and nlopt multi, respectively. For $m = 3$, the maximum power loss was $0.51\%$ and $0.0019\%$ (in absolute terms) for nlopt and nlopt multi, respectively.

\begin{table}[ht!]
\centering
\begin{tabular}{c|ccc|ccc}
\textbf{Method} & \multicolumn{3}{c|}{\textbf{Two treatments}} & \multicolumn{3}{c}{\textbf{Three treatments}}  \\ 
& $\approx$ weights & Euclidean distance & $\geq$ power & $\approx$ weights & Euclidean distance & $\geq$ power \\
\hline 
\textbf{nleq} & 100\% & 1.3 & 100\% & 100\% & 2.0 & 100\%\\ 
\textbf{nlopt} & 86.8\% & 20.1 & 84.3\% & 93.5\% & 7.9 & 90.3\% \\ 
\textbf{nlopt multi} & 87.3\% & 11.9 & 85.3\% & 99.9\% & 2.1 & 98.3\%\\ 
 
\end{tabular} 

\caption{Comparison of different methods compared with grid search, in terms of percentage of optimal weights which are within rounding error of the grid search weights (``$\approx$ equal" column), sum of Euclidean distance with the grid search weights, and percentage of realised disjunctive power that is equal or greater than the disjunctive power using the grid search weights (``$\geq$ column"). Results are based on 1000 sets of uniform random treatment effect vectors. For grid search, the grid size was set equal to 0.005. \label{tab:compare_grid_search}}

\end{table}

Finally, we compare the realised disjunctive power using nlopt and nlopt multi with that found using nleq in Table~\ref{tab:compare_nleq}. Note that nleq always had greater or equal disjunctive power compared with nlopt and nlopt multi. Table~\ref{tab:compare_nleq} shows that for a substantial proportion of treatment effect vectors, both nlopt and nlopt multi lead to suboptimal weights and a lower disjunctive power compared with nleq. This is particularly the case for nlopt and for smaller values of $m$.

Again, the absolute magnitude of the power differences is small. The maximum power loss for nlopt compared to using nleq was $0.0040\%$, $0.51\%$, $0.35\%$ and $0.28\%$ (in absolute terms) for $m = 2, 3, 4, 5$, respectively. The maximum power loss for nlopt multi compared to using nleq was $0.0012\%$, $0.0019\%$, $0.0099\%$ and $0.0013\%$ (in absolute terms) for $m = 2,3,4,5$, respectively.

\begin{table}[ht!]

\begin{tabular}{c|cccc}

\textbf{Method} & \textbf{Two treatments} & \textbf{Three treatments} & \textbf{Four treatments} & \textbf{Five treatments} \\ 
\hline 

\textbf{nlopt} & 74.3\% & 74.6\% & 66.4\%  & 59.9\% \\ 

\textbf{nlopt multi} & 76.2\% & 85.4\% & 85.9\% & 90.7\% \\ 

\end{tabular} 

\caption{Percentage of realised disjunctive powers that are equal to the disjunctive power using the nleq optimal weights. Note there were no cases were using the nleq resulted in a lower disjunctive power. Results are based on 1000 sets of uniform random treatment effect vectors.\label{tab:compare_nleq}}

\end{table}

\section{Additional simulation results}
\label{Asec:sim_results}

\subsection{Consistent treatment effects}

\subsubsection{Two treatments}

%

\begin{figure}[ht!]
\includegraphics[width = 0.7\linewidth]{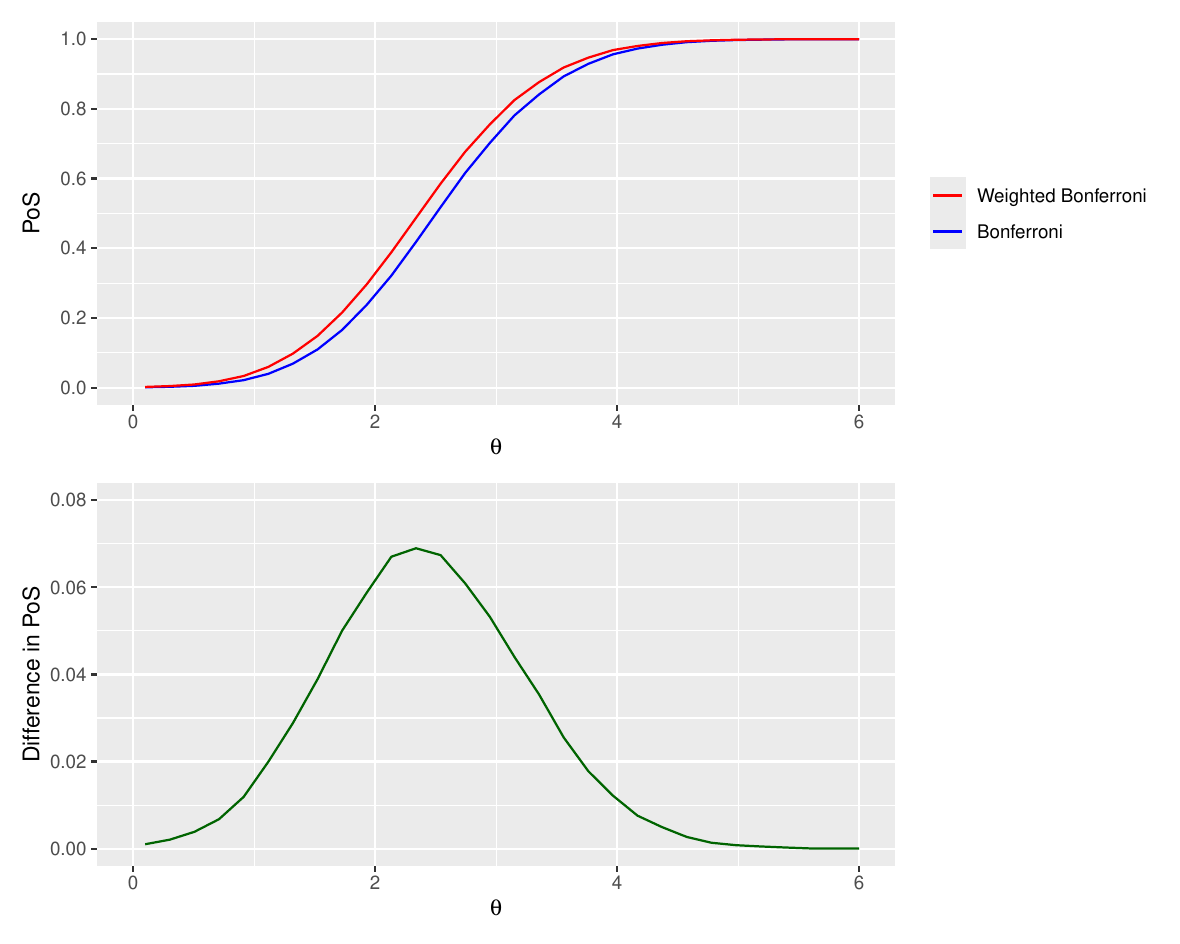} 
\centering
\caption{PoS for $H_2$ for $\bm{\theta} = (0, \theta)$. Results are from $10^5$ simulation replicates. \label{fig:PoS_2T_sim1}}
\end{figure}

Figure~\ref{fig:PoS_2T_sim1} shows the (marginal = disjunctive) PoS for $H_2$ under the scenario where $\bm{\theta} = (0, \theta)$.

%
%
%

Table~\ref{tab:rej_sim2a} shows the detailed results for rejection and non-rejection for the two hypothesis for $\theta=3$ based on 1000 simulations.

\begin{table}[ht!]
\scriptsize
\centering
\begin{minipage}{0.49\textwidth} 
\centering
\textbf{H1} \\[12pt] 
\begin{tabular}{c c | c c}
\multicolumn{2}{c|}{\textbf{Test}} & \multicolumn{2}{c}{\textbf{Weighted}} \\
 & & Rejection & No rejection \\
\hline
\multirow{2}{*}{\textbf{Unweighted}} & Rejection & 0  & 1 \\
& No rejection & 0 & 999
\end{tabular}
\end{minipage}
\hfill 
\begin{minipage}{0.49\textwidth} 
\centering
\textbf{H2} \\[12pt] 
\begin{tabular}{c c | c c}
\multicolumn{2}{c|}{\textbf{Test}} & \multicolumn{2}{c}{\textbf{Weighted}} \\
 & & Rejection & No rejection \\
\hline
\multirow{2}{*}{\textbf{Unweighted}} & Rejection & 715  & 0 \\
& No rejection & 50 & 235
\end{tabular}
\end{minipage}

\vspace{8pt}

\caption{\small Overall rejections of $H_1$ and $H_2$ when using the weighted and unweighted Bonferroni procedures for $\bm{\theta} = (0, 3)$. Results are from $1000$ simulation replicates. \label{tab:rej_sim2a}}

\end{table}

\begin{figure}[ht!]
\includegraphics[width = 0.7\linewidth]{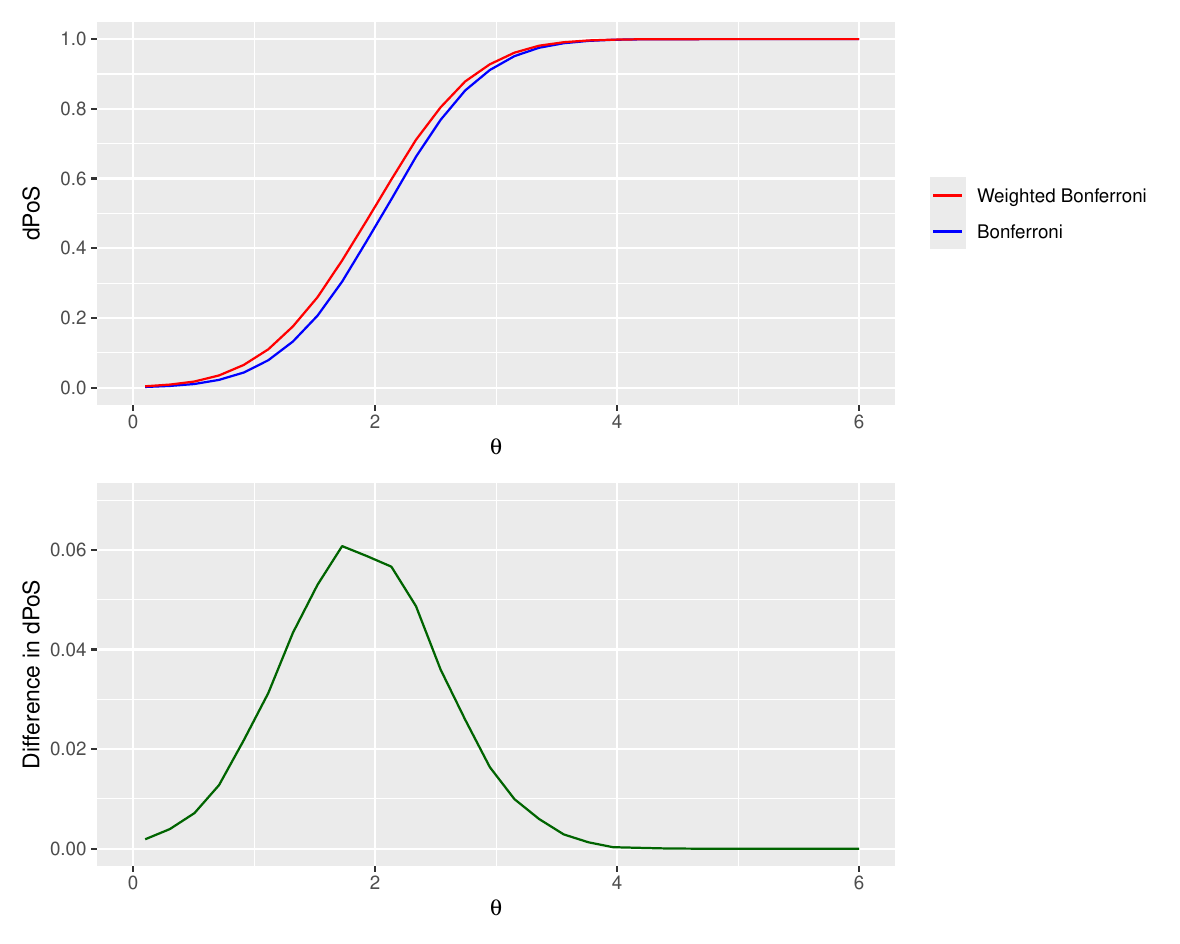} 
\centering
\caption{Disjunctive PoS (dPoS) for $\bm{\theta} = (\theta, \theta)$. Results are from $10^5$ simulation replicates. \label{fig:dPoS_2T_sim3}}
\end{figure}

\begin{figure}[ht!]
\includegraphics[width = \linewidth]{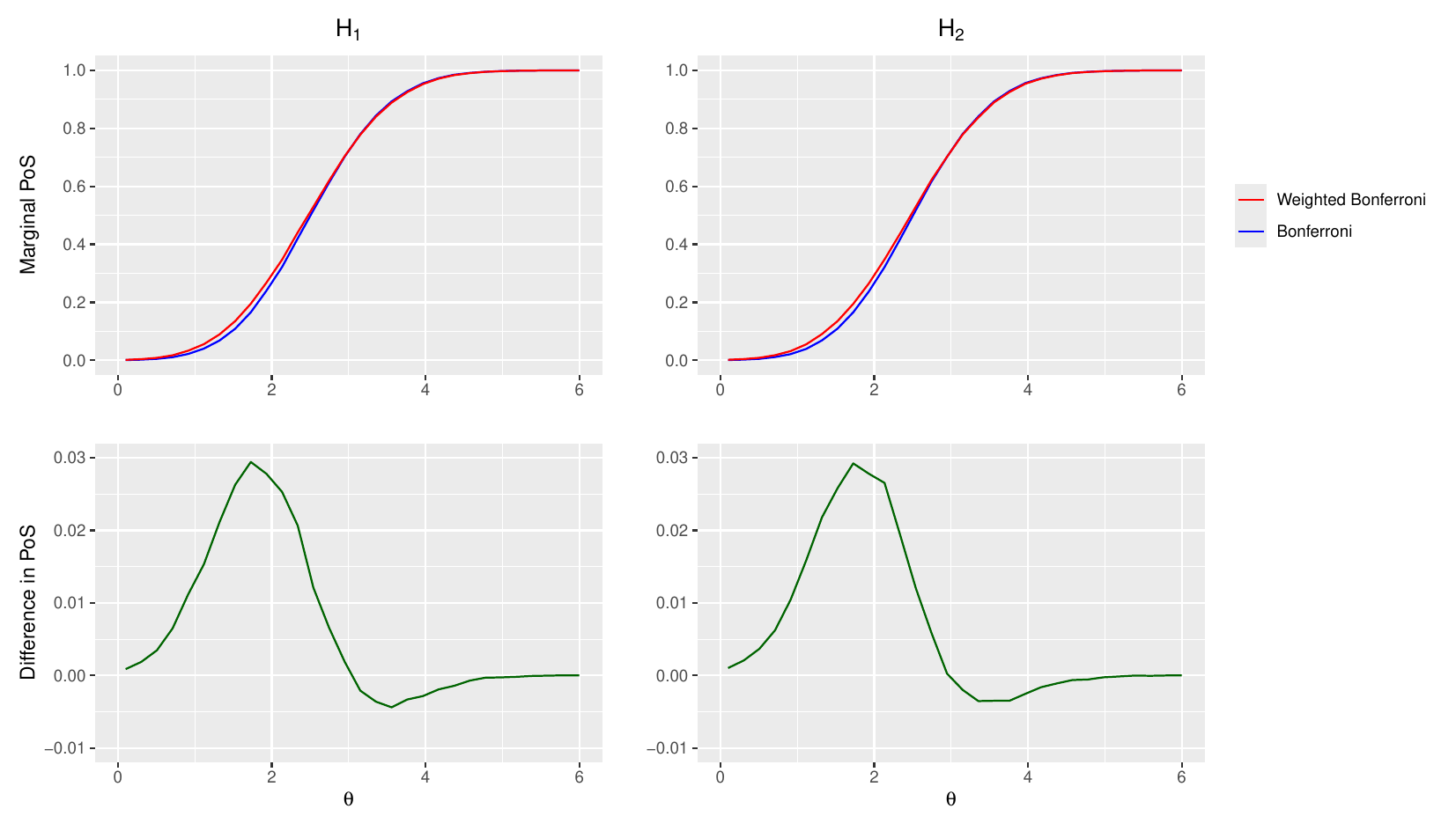} 
\centering
\caption{Marginal PoS (mPoS) for $H_1$ and $H_2$ when $\bm{\theta} = (\theta, \theta)$. Results are from $10^5$ simulation replicates.   \label{fig:mPoS_2T_sim3}}
\end{figure}

Now assume that the true standardised treatment means are given by $\bm{\theta} = (\theta, \theta)$, where $\theta > 0$.  Figure~\ref{fig:dPoS_2T_sim3} shows the disjunctive PoS while Figure~\ref{fig:mPoS_2T_sim3} shows the marginal PoS for $H_1$ and $H_2$.

%
%
%

Table~\ref{tab:rej_sim2c} shows the detailed results for rejection and non-rejection for the two hypothesis for $\theta=3$ based on 1000 simulations. 

\begin{table}[ht!]
\centering
\scriptsize
\begin{minipage}{0.49\textwidth} 
\centering
\textbf{H1} \\[12pt] 
\begin{tabular}{c c | c c}
\multicolumn{2}{c|}{\textbf{Test}} & \multicolumn{2}{c}{\textbf{Weighted}} \\
 & & Rejection & No rejection \\
\hline
\multirow{2}{*}{\textbf{Unweighted}} & Rejection & 741 & 0 \\
& No rejection & 35 & 224
\end{tabular}
\end{minipage}
\hfill 
\begin{minipage}{0.49\textwidth} 
\centering
\textbf{H2} \\[12pt] 
\begin{tabular}{c c | c c}
\multicolumn{2}{c|}{\textbf{Test}} & \multicolumn{2}{c}{\textbf{Weighted}} \\
 & & Rejection & No rejection \\
\hline
\multirow{2}{*}{\textbf{Unweighted}} & Rejection & 713  & 2 \\
& No rejection & 44 & 241
\end{tabular}
\end{minipage}

\vspace{8pt}

\caption{Overall rejections of $H_1$ and $H_2$ when using the weighted and unweighted Bonferroni procedures for $\bm{\theta} = (3, 3)$. Results are from $10^3$ simulation replicates. \label{tab:rej_sim2c}}

\end{table}

\clearpage
\subsubsection{$m>2$ treatments}

\begin{figure}[ht!]
\includegraphics[width = 0.7\linewidth]{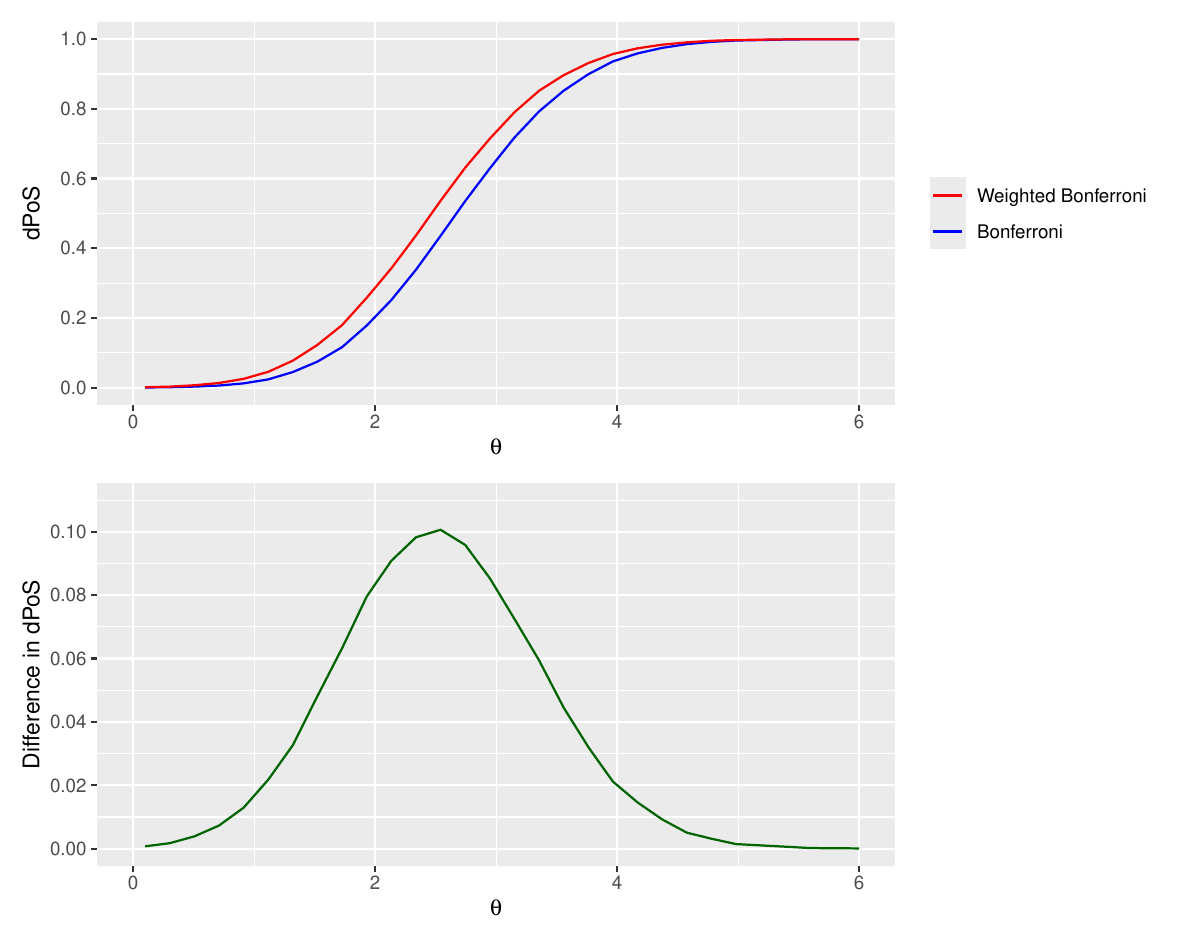} 
\centering
\caption{Probability of Success (PoS) for $H_3$ for $\bm{\theta} = (0, 0, \theta)$. Results are from $10^5$ simulation replicates. Positive values of the difference in PoS indicate where the PoS of the weighted Bonferroni approach is higher than the unweighted Bonferroni approach. \label{fig:dPoS_3T_sim1}}
\end{figure}

Figure~\ref{fig:dPoS_3T_sim1} shows the marginal PoS for $H_3$ when $\bm{\theta} = (0, 0, \theta)$.

\begin{figure}[ht!]
\includegraphics[width = 0.7\linewidth]{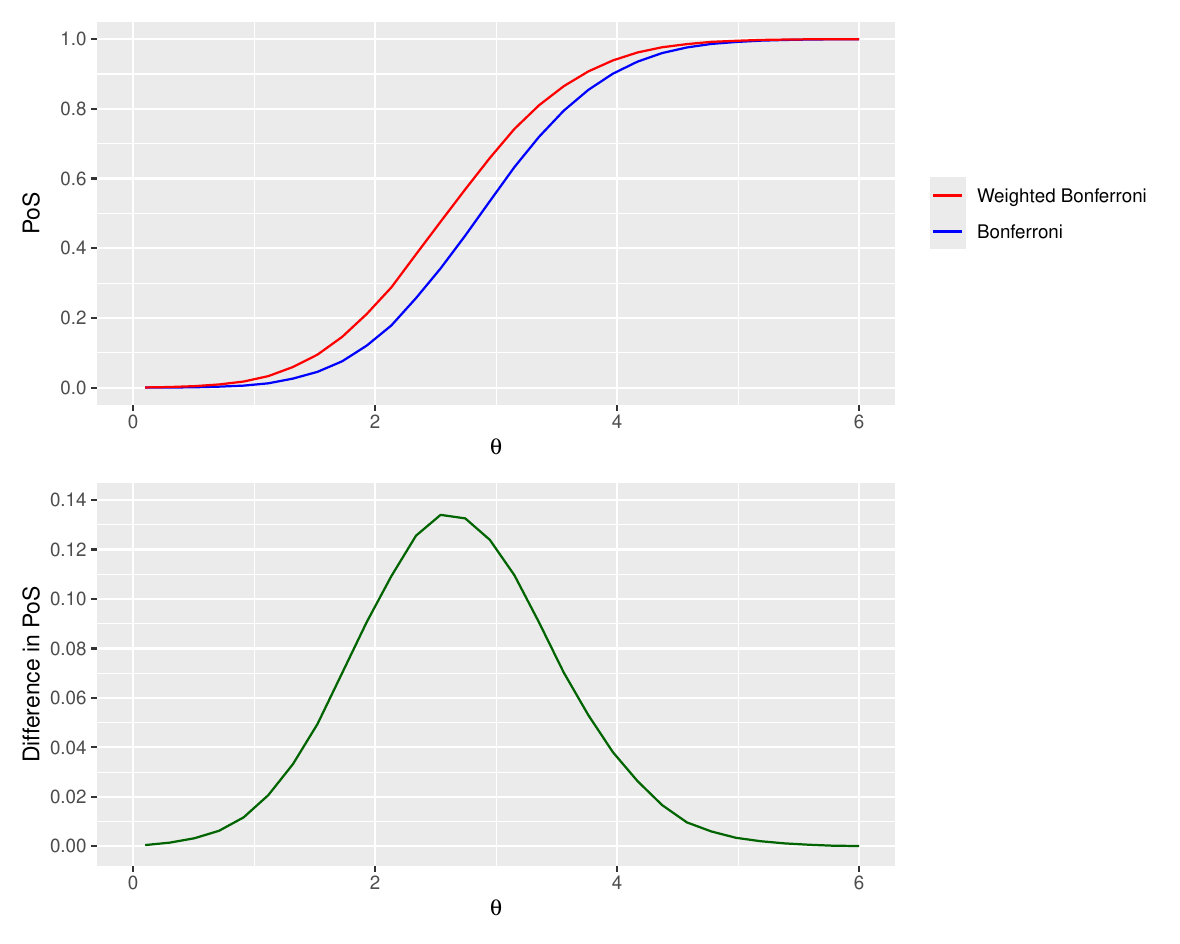} 
\centering
\caption{Probability of Success (PoS) for $H_5$ for $\bm{\theta} = (0, 0, 0, 0, \theta)$. Results are from $10^5$ simulation replicates.  \label{fig:mPoS_5T_sim1}}
\end{figure}

Figure~\ref{fig:mPoS_5T_sim1} shows the (marginal = disjunctive) PoS for $\bm{\theta} = (0, 0, 0, 0,\theta)$.

\begin{figure}[ht!]
\includegraphics[width = 0.7\linewidth]{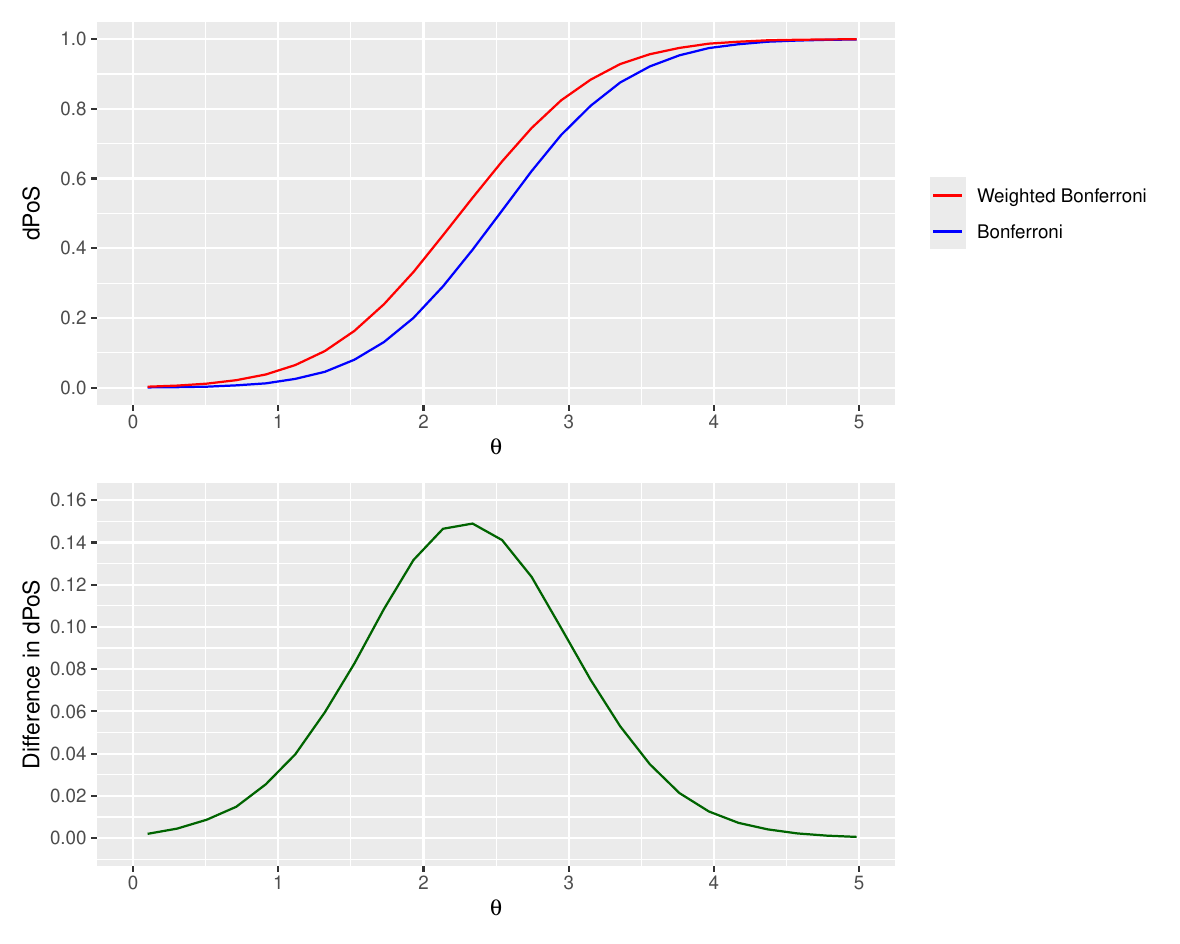} 
\centering
\caption{Disjunctive Probability of Success (dPoS) for $\bm{\theta} = (\theta_1, \theta_2, \theta_3, \theta_4,\theta)$, with $\theta_i \sim U[0, \theta]$ independently for $i = 1, 2, 3, 4$ and $\theta > 0$. Results are from $10^5$ simulation replicates. \label{fig:dPoS_5T_sim3}}
\end{figure}

\begin{figure}[ht!]
\includegraphics[width = 0.7\linewidth]{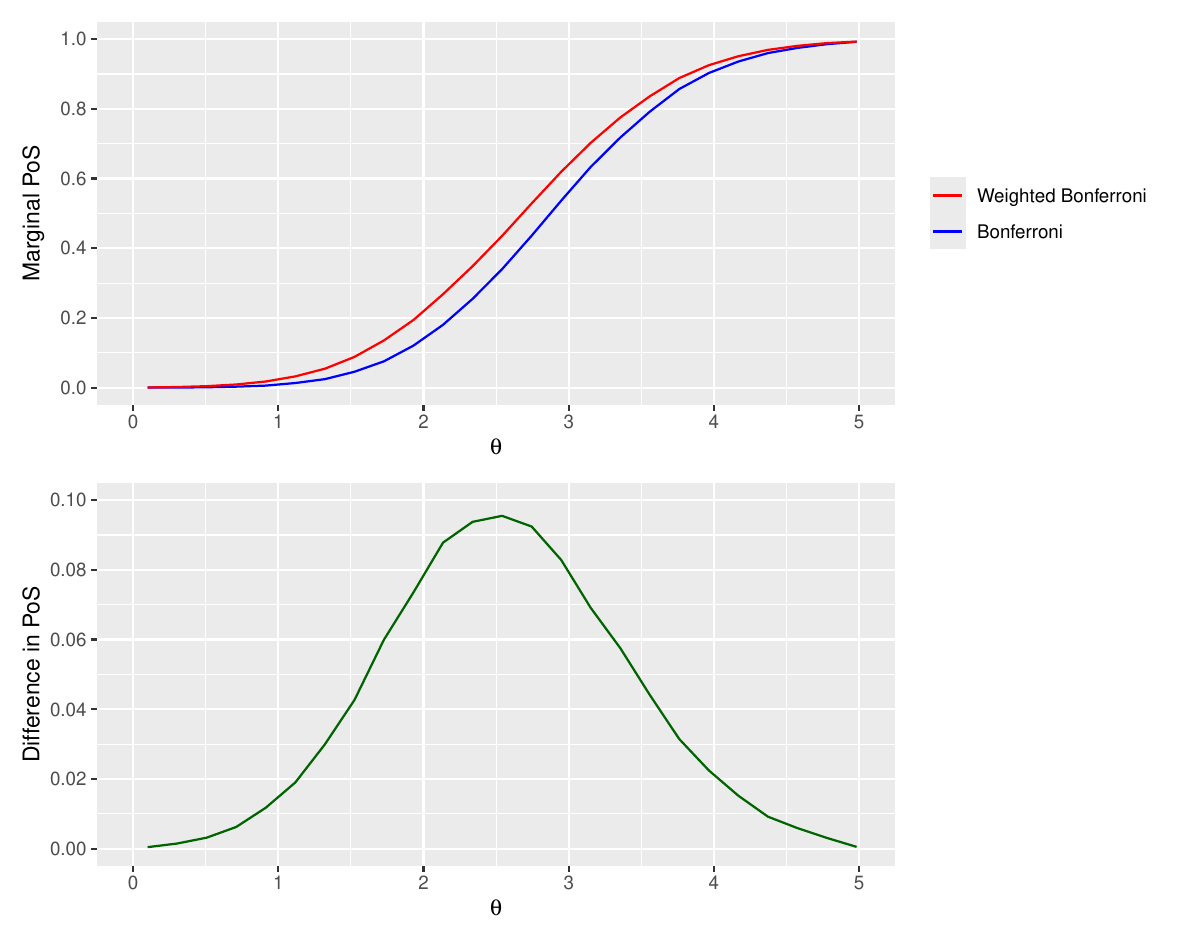} 
\centering
\caption{Marginal Probability of Success (PoS) for $H_5$ for $\bm{\theta} = (\theta_1, \theta_2, \theta_3, \theta_4,\theta)$, with $\theta_i \sim U[0, \theta]$ independently for $i = 1, 2, 3, 4$ and $\theta > 0$. Results are from $10^5$ simulation replicates. \label{fig:mPoS_5T_sim3}}
\end{figure}

Figure~\ref{fig:dPoS_5T_sim3} shows the disjunctive PoS for $\bm{\theta} = (\theta_1, \theta_2, \theta_3, \theta_4,\theta)$, with $\theta_i \sim U[0, \theta]$ independently for $i = 1, 2, 3, 4$ and $\theta > 0$, while Figure~\ref{fig:mPoS_5T_sim3} shows the marginal PoS for $H_5$.

\clearpage
\subsection{Robustness}

\begin{figure}[ht!]
\includegraphics[width = 0.7\linewidth]{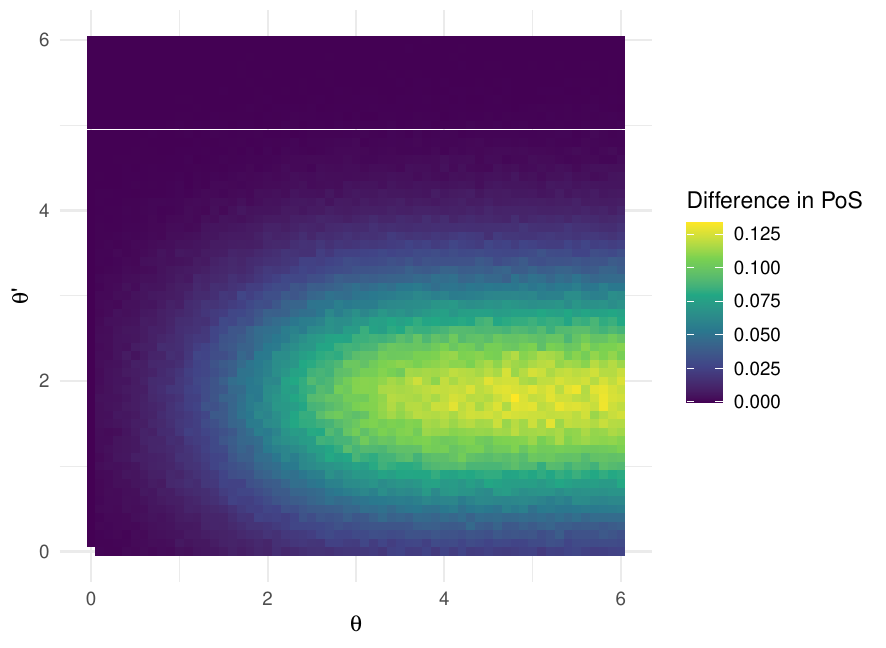} 
\centering
\caption{Heatmap showing difference in the Probability of Success (PoS) when trial~1 has treatment means given by $\bm{\theta} = (0, \theta)$ while trial~2 has treatment means given by $\bm{\theta'} = (0, \theta')$, where $0 \leq \theta, \theta' \leq 3$. Results are from $10^4$ simulations for each pair of values of $(\theta, \theta')$. Positive values indicate where the PoS of the weighted Bonferroni approach is higher than the PoS of the unweighted Bonferroni approach. \label{fig:dPoS_2T_heatmap_robust_a}}
\end{figure}

\begin{figure}[ht!]
\includegraphics[width = 0.7\linewidth]{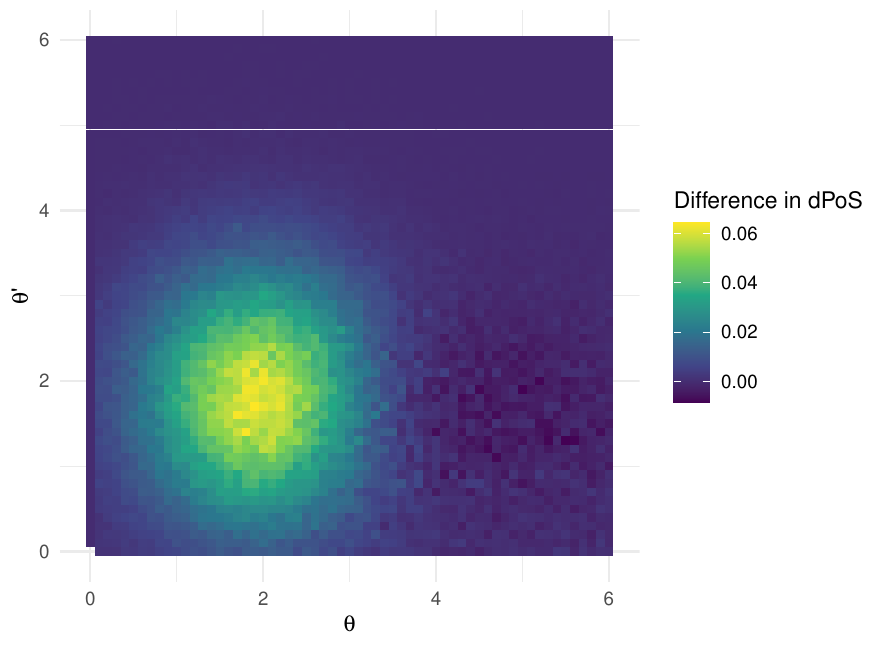} 
\centering
\caption{Heatmap showing difference in the disjunctive Probability of Success (dPoS) when trial~1 has treatment means given by $\bm{\theta} = (\theta, \theta)$ while trial~2 has treatment means given by $\bm{\theta'} = (\theta', \theta')$, where $0 \leq \theta, \theta' \leq 3$. Results are from $10^4$ simulations for each pair of values of $(\theta, \theta')$. Positive values indicate where the PoS of the weighted Bonferroni approach is higher than the PoS of the unweighted Bonferroni approach. \label{fig:dPoS_2T_heatmap_robust_c}}
\end{figure}

\end{document}